\documentclass[preprint2]{aastex}
\usepackage{amssymb}
\usepackage{amsmath}
\usepackage{comment}
\usepackage{fixltx2e} 
\newcommand\ghrs{GHRS}
\newcommand\fuse{\emph{FUSE}}
\newcommand\hst{\emph{HST}}
\newcommand\stis{STIS}
\providecommand{\cm[1]}{\ensuremath{\,{\rm cm}^{#1}}}
\providecommand{\kms}{\ensuremath{\,{\rm km\,s}^{-1}}}
\providecommand{\mA}{\ensuremath{\,\mbox{m\AA}}}
\providecommand{\Ang}{\ensuremath{\,\mbox{\AA}}}
\providecommand{\kpc}{\ensuremath{\,\mathrm{kpc}}}
\providecommand{\Lya}{\ensuremath{\mathrm{Ly}\alpha}} 
\providecommand{\Lyb}{\ensuremath{\mathrm{Ly}\beta}} 
\providecommand{\HH}{\ensuremath{\mathrm{H}_{2}}}
\providecommand{\Sithr}{\ensuremath{\mathrm{Si}^{+3}}}
\providecommand{\Cthr}{\ensuremath{\mathrm{C}^{+3}}}
\providecommand{\zabs}{\ensuremath{z_{\rm abs}}}
\providecommand{\zsiiv}{\ensuremath{z_{1393}}}
\providecommand{\zem}{\ensuremath{z_{\rm em}}}

\providecommand{\dvabs}{\ensuremath{\delta v_{\rm abs}}}

\providecommand{\EWr}{\ensuremath{W_{r}}}
\providecommand{\sigEWr}{\ensuremath{\sigma_{\EWr}}}
\providecommand{\EWlin[1]}{\ensuremath{\EWr{}_{,\mathrm{#1}}}}
\providecommand{\sigEWlin[1]}{\ensuremath{\sigma_{\EWr,\mathrm{#1}}}}
\providecommand{\EWo}{\ensuremath{W_{\rm obs}}}
\providecommand{\sigEWo}{\ensuremath{\sigma_{\EWo}}}

\providecommand{\wvr}{\ensuremath{\lambda_{r}}}

\providecommand{\wvlo}{\ensuremath{\lambda_{l}}}
\providecommand{\wvhi}{\ensuremath{\lambda_{h}}}
\providecommand{\N[1]}{\ensuremath{N_{\rm #1}}}
\providecommand{\NSiIV}{\ensuremath{N(\Sithr)}}
\providecommand{\NCIV}{\ensuremath{N(\Cthr)}}

\providecommand{\logSiIV}{\ensuremath{\log \NSiIV}}

\providecommand{\logN}{\ensuremath{\log N}}
\providecommand{\siglogN}{\ensuremath{\sigma_{\logN}}}
\providecommand{\IRatio}{\ensuremath{\NSiIV\slash\NCIV}}
\providecommand{\OmSiIV}{\ensuremath{\Omega_{\Sithr}}}
\providecommand{\OmCIV}{\ensuremath{\Omega_{\Cthr}}}

\providecommand{\ff[1]}{\ensuremath{f(#1)}}
\providecommand{\aff[1]}{\ensuremath{\alpha_{#1}}}
\providecommand{\kff[1]}{\ensuremath{k_{#1}}}
\providecommand{\Num}{\ensuremath{\mathcal{N}}}

\providecommand{\DXp}{\ensuremath{\Delta X}}

\providecommand{\DX[1]}{\ensuremath{\DXp(#1)}}
\providecommand{\ud}{\ensuremath{\mathrm{d}}}

\providecommand{\dNSiIVdz}{\ensuremath{\ud \Num_{\mathrm{Si\,IV}}/\ud z}}

\providecommand{\dNSiIVdX}{\ensuremath{\ud \Num_{\mathrm{Si\,IV}}/\ud X}}
\newcommand\ie{\emph{i.e.},\ }
\newcommand\eg{\emph{e.g.},\ }

\begin{document}


\title{The Last Eight-Billion Years of Intergalactic \ion{Si}{4} Evolution}

\author{Kathy L. Cooksey\altaffilmark{1}, J. Xavier
  Prochaska\altaffilmark{2} , Christopher Thom\altaffilmark{3}, and
  Hsiao-Wen Chen\altaffilmark{4}}

\altaffiltext{1}{NSF Astronomy \& Astrophysics Postdoctoral Fellow,
  MIT Kavli Institute for Astrophysics \& Space Research, 77 Massachusetts
    Avenue, 37-611, Cambridge, MA 02139, USA; kcooksey@space.mit.edu.}
\altaffiltext{2}{Department of Astronomy \& UCO/Lick Observatory,
  University of California, 1156 High Street, Santa Cruz, CA 95064, USA;
  xavier@ucolick.org}
\altaffiltext{3}{Space Telescope Science Institute, 3700 San Martin
  Drive, Baltimore, MD 21218, USA; cthom@stsci.edu.}  
\altaffiltext{4}{Department of Astronomy, University of Chicago, 5640
  South Ellis Avenue, Chicago, IL 60637, USA;
  hchen@oddjob.uchicago.edu.}

\shorttitle{Intergalactic \ion{Si}{4}} \shortauthors{Cooksey et al.}

\slugcomment{Draft 3: \today (Accepted to ApJ)} 


\begin{abstract}
  We identified 24 \ion{Si}{4} absorption systems with $z \lesssim 1$
  from a blind survey of 49 low-redshift quasars with archival {\it
    Hubble Space Telescope} ultraviolet spectra. We relied solely on
  the characteristic wavelength separation of the doublet to
  automatically detect candidates. After visual inspection, we defined
  a sample of 20 definite (group G = 1) and 4 ``highly-likely'' (G =
  2) doublets with rest equivalent widths \EWr\ for both lines
  detected at $\ge 3\sigEWr$. The absorber line density of the G = 1
  doublets was $\dNSiIVdX = 1.4^{+0.4}_{-0.3}$ for $\logSiIV >
  12.9$. The best-fit power law to the G = 1 frequency distribution of
  column densities \ff{\NSiIV}\ had normalization $k =
  (1.2^{+0.5}_{-0.4}) \times 10^{-14}\cm{2}$ and slope $\aff{N} =
  -1.6^{+0.3}_{-0.3}$. Using the power-law model of \ff{\NSiIV}, we
  measured the \Sithr\ mass density relative to the critical density:
  $\OmSiIV = (3.7^{+2.8}_{-1.7}) \times 10^{-8}$ for $13 \le \logSiIV
  \le 15$. From Monte Carlo sampling of the distributions, we
  estimated our value to be a factor of $4.8^{+3.0}_{-1.9}$ higher
  than the $2 \le z \le 4.5$ $\langle \OmSiIV \rangle$. From a simple
  linear fit to \OmSiIV\ over the age of the Universe, we estimated a
  slow and steady increase from $z = 5.5 \rightarrow 0$ with $\ud
  \OmSiIV \slash \ud t_{\rm age} = (0.61\pm0.23) \times 10^{-8}\,{\rm
    Gyr}^{-1}$. We compared our ionic ratios \IRatio\ to a $2 < z <
  4.5$ sample and concluded, from survival analysis, that the two
  populations are similar, with median $\langle\IRatio \rangle =
  0.16$.
\end{abstract}



\section{Introduction}\label{sec.intro}

\addtocounter{footnote}{4}

The signatures of the cosmic enrichment cycle are etched into the
processed gas and reflected in its metallicity, elemental abundances,
density, and/or spatial distribution.  Measuring these quantities
constrain models of galactic feedback processes \citep[see][and
references therein]{bertschinger98}. Currently, quasar absorption-line
(QAL) spectroscopy is the best tool for probing the IGM. 

The ultraviolet (UV) transition \ion{Si}{4} $\lambda\lambda 1393.76$,
1402.77\Ang\ is a well-studied doublet in QAL surveys covering $1.5 <
z < 5.5$. From the observability perspective, the \ion{Si}{4}
absorption lines are valuable for the following three reasons. First,
\ion{Si}{4} absorbers can be observed outside the \Lya\ forest since
they have rest wavelengths red-ward of \Lya\ $\lambda1215$. This
reduces the effect of blending. Second, they are observable from
ground-based telescopes when they redshift into the optical passband
at $z > 1.5$.  Third, they constitute a doublet with characteristic
rest wavelength separation ($9\Ang$) and equivalent width ratio
($2:1$, respectively, in the unsaturated regime). When these
distinctive criteria are met, we can be fairly confident that the pair
of absorption lines are a \ion{Si}{4} doublet.

From the astrophysics perspective, the \ion{Si}{4} doublet is a
strong, observable transition of silicon. The abundance of silicon is
predominately driven by Type II supernovae at $z \gtrsim 1$, with an
increasing fraction from feedback from asymptotic giant branch stars
more recently \citep{oppenheimeranddave08, wiersmaetal09b}. Thus, at $z
\gtrsim 1$, silicon traces oxygen, the most abundant metal. Oxygen
itself is difficult to study since its strong transitions are blue-ward
of \Lya\ (\eg \ion{O}{6} $\lambda\lambda 1031,1037$), if not also
blue-ward of the Lyman limit (\eg \ion{O}{4} $\lambda787$). By using
\ion{Si}{4} absorption as a tracer of oxygen, \citet{songaila01}
constrained the IGM metallicity to be $>\!10^{-3.5}\,Z_{\odot}$ at $z
= 5$. Therefore, the fraction of cosmic star formation that occurred
before $z = 5$, or within $1\,$Gyr of the Big Bang, was
$>\!10^{-3.5}$.

\citet{songaila01} also measured \OmSiIV, the \Sithr\ mass density
relative to the critical density, for $2 < z < 5.5$. The \Sithr\ mass
density may represent the largest contribution to the silicon mass
density $\Omega_{\rm Si}$ at these redshifts \citep{songaila01} but
still constitute only a small fraction of $\Omega_{\rm Si}$
\citep{aguirreetal04}. For $z = 4.5 \rightarrow 2$, \OmSiIV\ was
roughly constant at $\approx\!1.2\times 10^{-8}$ for absorbers with
column densities $13 \le \logSiIV \le 15$.\footnote{We adjust
  quantities from other studies to our adopted cosmology: $H_{0} =
  70\kms\,{\rm Mpc}^{-1}$, $\Omega_{\rm M} = 0.3$, and
  $\Omega_{\Lambda}=0.7$.} For $z = 5.5 \rightarrow 4.5$, the \Sithr\
mass density may have increased by an order of magnitude.  Subsequent
studies have largely supported these broad trends in \OmSiIV\
(non)evolution \citep{boksenbergetal03ph, songaila05,
  scannapiecoetal06}.

Observations of gas bearing both \ion{Si}{4} and \ion{C}{4}
$\lambda\lambda 1548,1550\Ang$ absorption offer constraints on the
shape, spatial extent, and/or evolution of the ionizing ultraviolet
background \citep[UVB;][]{boksenbergetal03ph, aguirreetal04,
  scannapiecoetal06}. The ionization threshold for ${\rm
  Si}^{++}$-to-\Sithr\ is $2.5\,$Ryd and for ${\rm C}^{++}$-to-\Cthr,
$3.5\,$Ryd. Thus, the ionic ratio \IRatio\ is affected by the shape of
the UVB at these energies, and the fluctuations in the ratio spatially
and in time could constrain the patchiness and evolution, respectively,
of the UVB.

Of particular interest is the effect of \ion{He}{2} reionization on
the UVB at $z \approx 3$, which is expected to boost \ion{Si}{4}
absorption and suppress \ion{C}{4} absorption
\citep{madauandhaardt09}. Several studies find no evidence for a sharp
break in the shape of the UVB at $z \approx 3$ or even significant
evolution in its shape for $z \approx 4.5 \rightarrow 1.5$ from
studies of \ion{Si}{4} and \ion{C}{4} systems with column densities of
$10^{12}\cm{-2}$ to $10^{14}\cm{-2}$ \citep{kimetal02,
  boksenbergetal03ph, aguirreetal04}. Indeed, any variation in the
ionic ratio \IRatio\ may be dominated by the variation in the
metallicity of the absorbing gas \citep{boltonandviel10ph}. However,
evidence for both a break and strong evolution in the UVB have been
detected in some studies of \ion{Si}{4} and \ion{C}{4} absorbers
\citep{songaila98, songaila05}.

The gas giving rise to $12 < \logSiIV < 14$ absorption has $T \lesssim
10^{4.9}\,$K and nearly constant ${\rm [Si/C]} \approx +0.77$ for $1.5
< z < 4.5$ \citep{aguirreetal04}.\footnote{We adopt the following
  notation: ${\rm [Si/C]} = \log (n_{\rm Si}/n_{\rm C}) - \log
  (n_{{\rm Si},\odot}/n_{{\rm C},\odot})$, where $n_{\rm X}$ is the
  volume density of element X.} In simulations of the IGM, most
(possibly all) of the silicon is located in distinct clouds of
metal-enriched gas.  At $z \approx 3$, the clouds have radii
$\approx\!1\,$Mpc \citep[proper;][]{scannapiecoetal06} and could be
considered filamentary structure. However, by $z = 0$, the enriched
clouds are actually the extended gaseous halos ($\approx\!100\kpc$) of
galaxies \citep{daveandoppenheimer07}.  Absorbers in the
circum-galactic medium would likely be subjected to a softer ionizing
background (due to the increased stellar contribution) than the
\citet{haardtandmadau96} UVB typically used in IGM studies. Local
sources (\ie star-forming galaxies with a non-zero escape fraction of
ionizing photons) may be the most important contributor to the
background. If the background for \ion{Si}{4} absorbers were softer at
$z \lesssim 1$, [Si/C] would be lower \citep{aguirreetal04}.

The current work finalizes our analysis of archival {\it Hubble Space
  Telescope} (\hst) UV spectra, gathered prior to Servicing Mission 4
(UT July 2009).  This is the largest survey for \ion{Si}{4} systems at
$z \lesssim 1$ to date and covers the last eight-billion years of the
cosmic enrichment cycle. In other words, the net effect of cosmic star
formation (and feedback) at the `end' (\ie $z = 0$) is constrained by
observations of \ion{Si}{4} absorption at low redshift. Also, the
survey of $z \lesssim 1$ \ion{Si}{4} absorbers provides a baseline for
similar, high-redshift surveys, which are currently more numerous. The
data reduction and analysis methods used in this paper are described
in detail in \citet[][hereafter Paper
I]{cookseyetal10},\defcitealias{cookseyetal10}{Paper I} to which the
interested reader is referred.

This paper is organized as follows: we present the data processing and
sample selection in Sections \ref{sec.data} and \ref{sec.selec},
respectively; we analyze and discuss the frequency distribution, its
moments, and \IRatio\ in Section \ref{sec.analysis}; and Section
\ref{sec.summ} is a summary.


\section{Data, Reduction, and Measurements}\label{sec.data}

\begin{figure}[!hbt]
  \begin{center}
  \includegraphics[height=0.47\textwidth,angle=90]{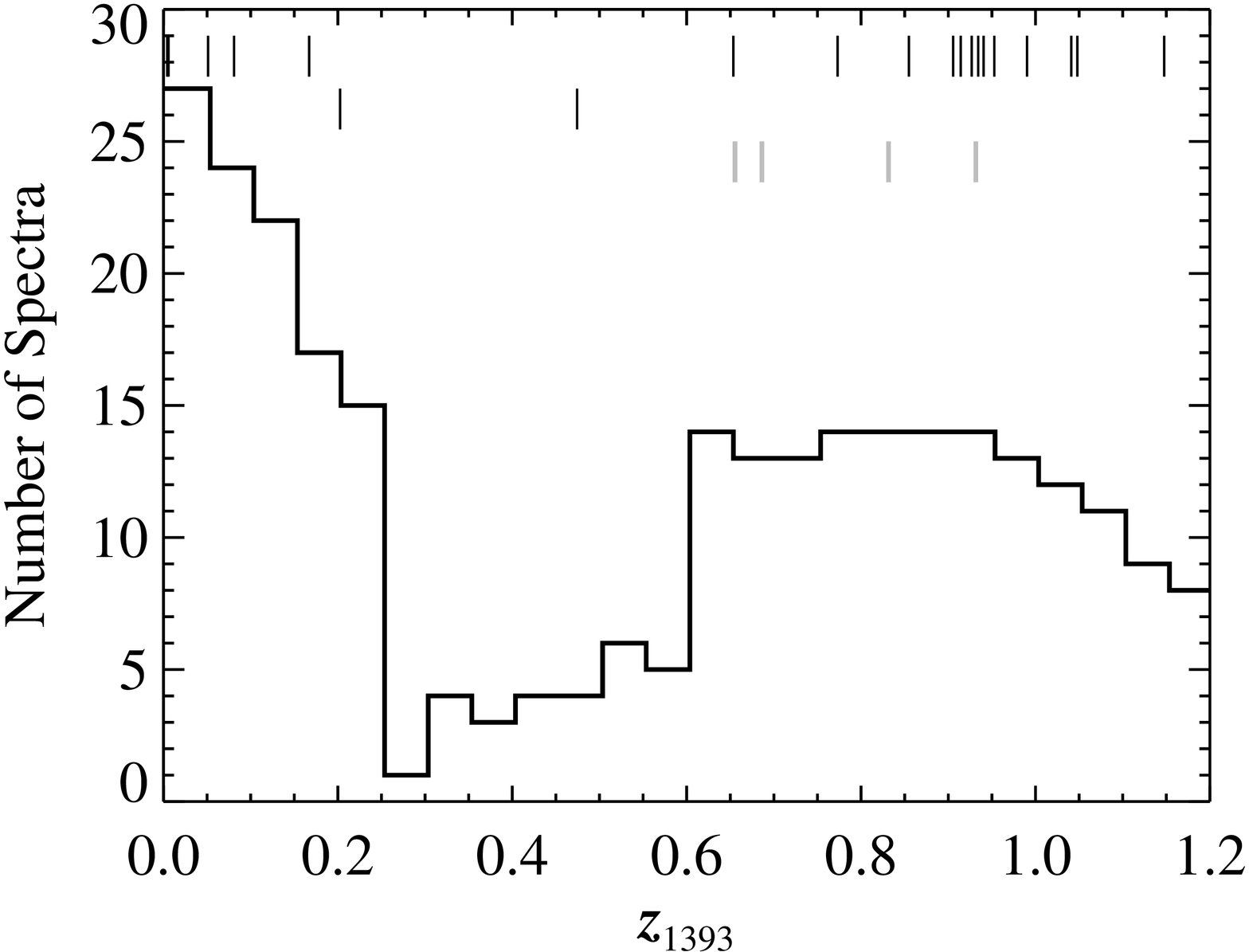} 
  \end{center}
  \caption[Schematic of redshift coverage for the current survey.]
  {Schematic of redshift coverage for the current survey. The number
    of spectra with coverage of the \ion{Si}{4} doublet is shown as a
    function of \zsiiv\ (histogram). The \stis\ E140M spectra covered
    $\zsiiv \lesssim 0.21$. The \stis\ E230M spectra typically covered
    $0.6 \lesssim \zsiiv < 1.2$.  The redshift range $0.21 \lesssim
    \zsiiv \lesssim 0.6$ was covered by some E230M spectra as well as
    \stis\ medium-resolution gratings and \ghrs\ spectra. The
    redshifts of the doublets detected with $\EWr \ge 3\,\sigEWr$ in
    both lines are shown with the hashes across the top. The top and
    middle rows indicate the redshifts of the 18 unsaturated and
    two saturated doublets, respectively, in the definite group (G = 1). The
    bottom row shows the redshift of the four unsaturated
    ``highly-likely'' (G = 2) doublets.
    \label{fig.gz}
  }
\end{figure}

We conducted a blind survey for $z \lesssim 1$ \ion{Si}{4} systems in
the {\it Hubble Space Telescope} (\hst) UV spectra of 49 low-redshift
quasars, which makes the current work the largest low-redshift
\ion{Si}{4} study to date. We included spectra from the Space
Telescope Imaging Spectrograph (\stis; pre-Servicing Mission 4) and
the Goddard High-Resolution Spectrograph (\ghrs). The \stis\ echelle
spectra, taken with the E230M grating, provided most of the search
path length (see Figure \ref{fig.gz}), but we also searched the other
\stis\ echelle grating (E140M) and the \ghrs\ echelle (ECH-B) and
long-slit (G160M, G200M, G270M) gratings. Spectra from the {\it Far
  Ultraviolet Spectroscopic Explorer} (\fuse) covered the transitions
with rest wavelengths $\wvr < 1100\Ang$ (\eg higher-order \ion{H}{1}
Lyman lines). All spectra had resolution with full-width at
half-maximum ${\rm FWHM} \le 15\kms$ and signal-to-noise ratio ${\rm
  S/N} \ge 2\,{\rm pix}^{-1}$.

The spectra were retrieved from the Multimission Archive at Space
Telescope (MAST).\footnote{See http://archive.stsci.edu/.} The
reduction and co-addition of multiple observations followed the
algorithms described in \citet{cookseyetal08}. The spectra were
normalized semi-automatically. All reduced, co-added, and normalized
spectra are available online, even those not explicitly searched in
this paper.\footnote{See http://www.ucolick.org/$\sim$xavier/HSTSiIV/
  for the normalized spectra, the continuum fits, the \ion{Si}{4}
  candidate lists, and the Monte Carlo completeness limits for all
  sightlines as well as the completeness test results for the full
  data sample.}

The central wavelength (and redshift) of an absorption line was
measured by the optical depth-weighted mean of the pixels within the
wavelength bounds ($\wvlo, \wvhi$) defining the absorption line (see
Table \ref{tab.cand}). The rest equivalent widths
$\EWr$ were measured with simple boxcar summation, and the column
densities (\eg \NSiIV) were measured with the apparent optical depth
method \citep[AODM;][]{savageandsembach91}. The doublet column density
\NSiIV\ is either: the variance-weighted mean of the measurements of
both lines; the column density from the one line with a measurement;
the greater lower limit; or the mean, when the limits of the lines
define a finite range.

\begin{figure}[!hbt]
  \begin{center}$
    \begin{array}{c}
      \includegraphics[height=0.45\textwidth,angle=90]{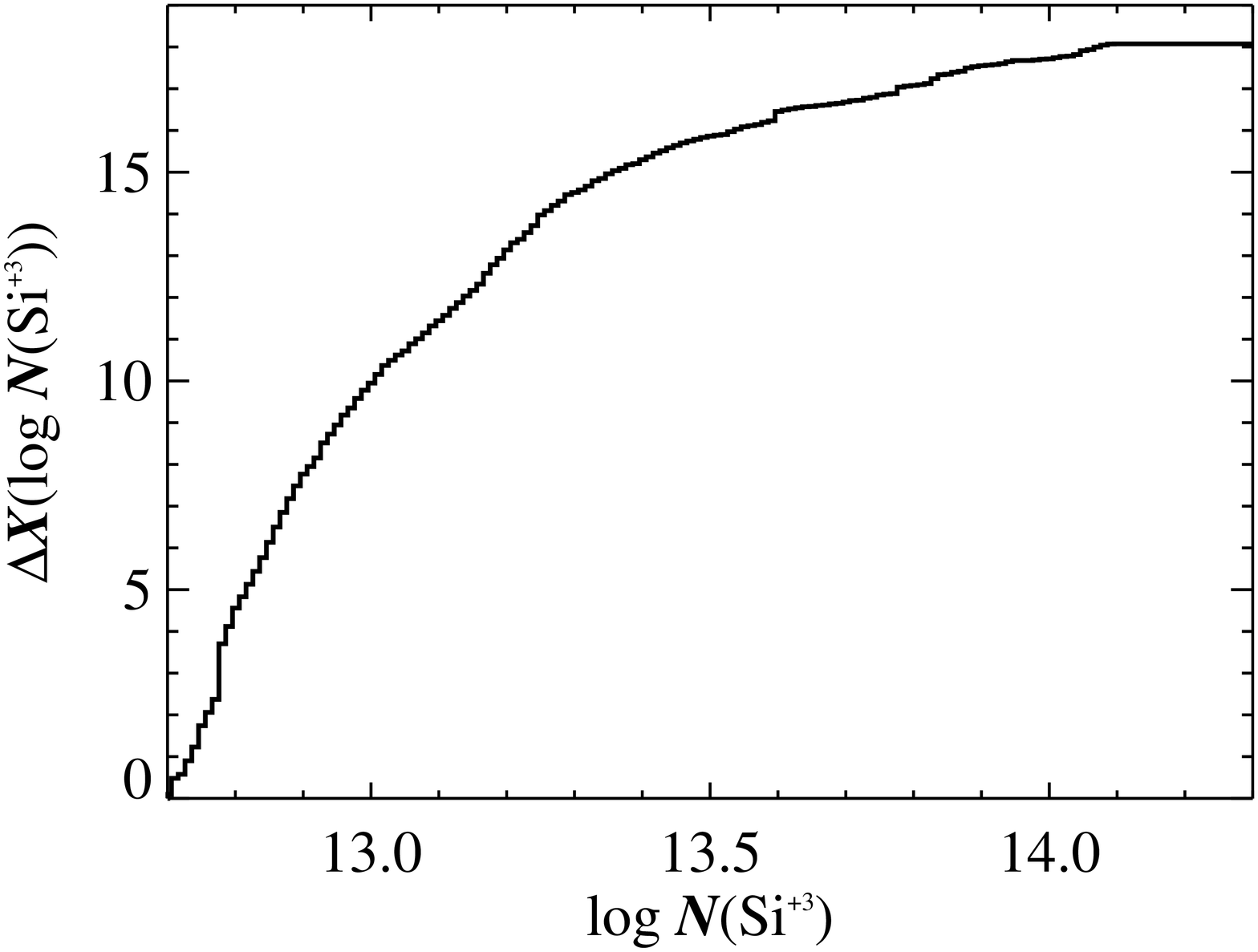} \\
      \includegraphics[height=0.45\textwidth,angle=90]{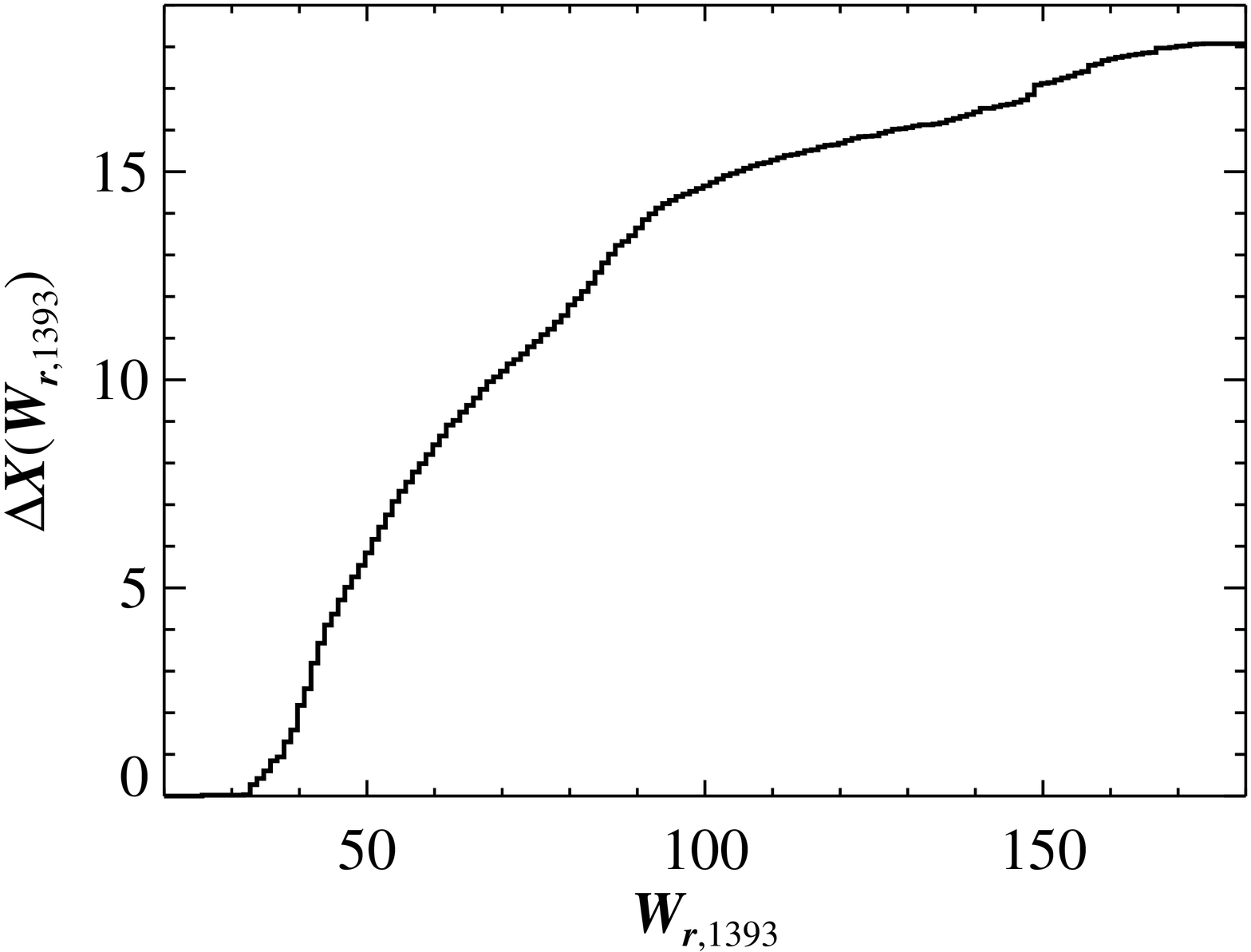}
    \end{array}$
  \end{center} 
  \caption[Redshift path length \DX{\logSiIV} and \DX{\EWlin{1393}}.]
  {Redshift path length \DX{\logSiIV} and \DX{\EWlin{1393}}\ as a
    function of \ion{Si}{4} 1393 column density (top) and rest equivalent
    width respectively (bottom). These estimates are based on Monte
    Carlo analysis and correspond to 95\% completeness limits.
    \label{fig.x}
  }
\end{figure}

The co-moving path length \DXp\ sensitive to \ion{Si}{4} doublets with
$\EWr \ge 3\sigEWr$ in both lines was estimated from Monte Carlo
simulations. For these simulations, we replaced all
automatically-detected features in the archival spectra with random
noise drawn from surrounding pixels. For each redshift bin ($\delta z
= 0.005$) of each spectrum, we distributed $10^3$ \ion{Si}{4} doublets
with a range of column densities, Doppler parameters, and number of
components. Then we measured the \NSiIV\ and \EWlin{1393}\ limit per
bin at which 95\% of the doublets were automatically recovered (see
Figure \ref{fig.x}). The available path length increases with
increasing \NSiIV\ and \EWlin{1393}; for the strongest absorbers
($\logSiIV \ge 14.1$ and $\EWlin{1393} \ge 174\mA$), $\DXp = 18$.

\begin{deluxetable}{rlllllllll}
\tablewidth{0pc}
\tablecaption{\ion{Si}{4} CANDIDATES SUMMARY \label{tab.cand}}
\tabletypesize{\scriptsize}
\tablehead{ 
\colhead{(1)} & \colhead{(2)} & \colhead{(3)} & \colhead{(4)} & \colhead{(5)} & 
\colhead{(6)} & \colhead{(7)} & \colhead{(8)} & \colhead{(9)} & \colhead{(10)} \\ 
\colhead{$z_{1393}$} & \colhead{\dvabs} & 
\colhead{$\lambda_{r}$} & \colhead{\wvlo} & 
\colhead{\wvhi} & \colhead{\EWr} & \colhead{\sigEWr} & 
\colhead{\logN} & \colhead{\siglogN} & 
\colhead{Flag} \\
 & \colhead{(\!\kms)} & \colhead{(\AA)} & \colhead{(\AA)} & 
\colhead{(\AA)} & \colhead{(\!\mA)} & \colhead{(\!\mA)} & & & 
}
\startdata
\hline \\[-1ex]
\multicolumn{10}{c}{MRK335 ($\zem=0.026$)} \\[1ex]
\hline
$-0.00001$ & \nodata & 1393.76 & 1393.40 & 1394.04 &     212 &      15 & 13.55 &  0.04 & 495 \\
 & $   -6.9$ & 1402.77 & 1402.41 & 1403.06 &     129 &      15 & 13.54 &  0.05 &  \\
$ 0.00643$ & \nodata & 1393.76 & 1402.41 & 1403.03 &     125 &      14 & 13.23 &  0.05 & 317 \\
 & $   -7.3$ & 1402.77 & 1411.48 & 1412.10 & $<     32$ & \nodata & $< 13.15$\tablenotemark{a} & \nodata &  \\
$ 0.01078$ & \nodata & 1393.76 & 1408.67 & 1408.91 & $<     20$ & \nodata & $< 12.65$\tablenotemark{a} & \nodata & 173 \\
 & $    0.7$ & 1402.77 & 1417.78 & 1418.02 &      37 &      12 & 12.97 &  0.14 &  \\
\hline \\[-1ex]
\multicolumn{10}{c}{PG0117+213 ($\zem=1.493$)} \\[1ex]
\hline
$ 0.63473$ & \nodata & 1393.76 & 2278.01 & 2278.98 &     622 &      70 & $> 13.84$\tablenotemark{a} & \nodata & 258 \\
 & $   10.0$ & 1402.77 & 2292.74 & 2293.72 & $<     83$ & \nodata & $< 13.59$  & \nodata &  \\
$ 0.63639$ & \nodata & 1393.76 & 2280.25 & 2281.14 & $<    116$ & \nodata & $< 13.41$\tablenotemark{a} & \nodata & 160 \\
 & $   -0.5$ & 1402.77 & 2295.00 & 2295.90 &     240 &      41 & $> 13.72$\tablenotemark{a} & \nodata &  \\
$ 0.63962$ & \nodata & 1393.76 & 2284.12 & 2286.44 &     478 &      67 & $> 13.95$  & \nodata & 259 \\
 & $   94.0$ & 1402.77 & 2298.89 & 2301.23 & $<    167$ & \nodata & $< 13.81$  & \nodata &  \\
\hline \\[-1ex]
\multicolumn{10}{c}{TONS210 ($\zem=0.116$)} \\[1ex]
\hline
$-0.00070$ & \nodata & 1393.76 & 1392.56 & 1393.09 &      81 &      12 & 13.04 &  0.06 & 367 \\
 & $   -4.4$ & 1402.77 & 1401.57 & 1402.10 & $<     25$ & \nodata & $< 12.77$  & \nodata &  \\
$-0.00002$ & \nodata & 1393.76 & 1393.34 & 1394.13 &     272 &      14 & 13.63 &  0.03 & 431 \\
 & $    7.8$ & 1402.77 & 1402.35 & 1403.15 &     110 &      17 & 13.47 &  0.06 &  \\
$ 0.00129$ & \nodata & 1393.76 & 1395.39 & 1395.69 &      29 &       9 & 12.55 &  0.14 & 301 \\
 & $   -8.0$ & 1402.77 & 1404.42 & 1404.72 & $<     26$ & \nodata & $< 12.79$  & \nodata &  \\
\enddata
\tablenotetext{a}{\logN\ measured by assuming \EWr\ results from the linear portion of the COG.}
\tablecomments{
Summary of Si\,IV doublet candidates by target and redshift of Si\,IV 1393.
Upper limits are $2\sigma$ limits for both \EWr\ and \logN.
The binary flag is described in Section \ref{sec.selec}.
\\
(This table is available in a machine-readable form in the online journal.  A portion is shown here for guidance regarding its form and content.)
}
\end{deluxetable}

\section{\ion{Si}{4} Sample Selection}\label{sec.selec}

Absorption lines with observed equivalent width $\EWo \ge 3\sigEWo$
were automatically detected in the spectra. Candidate \ion{Si}{4}
doublets were identified based solely on the characteristic wavelength
separation of the doublet ($\approx\!1940\kms$). First, each $\EWo \ge
3\sigEWo$ line was assumed to be \ion{Si}{4} 1393, and any
automatically-detected line that was near the location of would-be
\ion{Si}{4} 1402 was adopted as such. If no $\EWo \ge 3\sigEWo$ line
existed, an upper limit was set on \EWo\ from the spectrum. Second,
any automatically-detected line not already tagged as \ion{Si}{4} 1393
or 1402 was assumed to be the latter, and an upper limit was set on
$\EWo{}_{,1393}$ from the spectrum \citepalias[see][for
details]{cookseyetal10}. A summary of all candidate \ion{Si}{4}
doublets are given in Table \ref{tab.cand}.

Other common absorption lines (\eg \Lya\ and \ion{C}{4}) were
associated with the candidate systems in a similar manner. For
example, an automatically-detected absorption line would be identified
as a candidate \Lya\ line if it had observed wavelength $\lambda
\approx \lambda_{\alpha}(1+z_{\rm cand})$, where $\lambda_{\alpha}$ is
the rest wavelength of \Lya\ and $z_{\rm cand}$ is the redshift of the
candidate \ion{Si}{4} doublet. The \fuse\ spectra were useful in this
step, since they covered the \Lyb\ $\lambda1025$, \ion{O}{6}, and
\ion{C}{3} $\lambda977$ lines for $z \lesssim 0.1$ candidate systems.
The presence of associated lines increased the confidence of our
identifications.

Each candidate doublet was assigned a machine-generated, binary flag
that scaled with the number of desired characteristics of a true \ion{Si}{4}
doublet. The characteristics (flags) are as follows: 
\begin{itemize}
\item[256:] {$\EWlin{1393} \ge 3\sigEWlin{1393}$;}
\item[128:] {$\EWlin{1402} \ge 3\sigEWlin{1402}$;} 
\item[64:] {the ratio $\EWlin{1393}:\EWlin{1402}$ is in the range of
    $1:1$ to $2:1$ plus\slash minus the propagated error of the ratio;}
\item[32:] {the optical depth-weighted centroids of the \ion{Si}{4}
    lines have $|\dvabs| \le 10\kms$;\footnote{The wavelength
      separation between a line at \zabs\ and a \ion{Si}{4} line at
      \zsiiv\ is $\dvabs \equiv c(\zabs- \zsiiv)/ (1+\zsiiv)$.}} 
\item[16:] {there exists a candidate \Lya\ line with $\EWlin{\alpha}
    \ge 3\sigEWlin{\alpha}$;} 
\item[8:] {the 1393 line is outside of the \Lya\ forest}
\item[4:] {and outside the \HH\ forest;}
\item[2:] {the smoothed AOD per pixel of the doublet lines agree
    within $1\sigma$ for $\ge 68.3\%$ of the pixel; and}  
\item[1:] {there exists one or more candidate lines (not
    \ion{H}{1}) associated with the candidate doublet and with $\EWr
    \ge 3\sigEWr$.} 
\end{itemize}

\begin{figure}[!hbt]
  \includegraphics[width=0.45\textwidth]{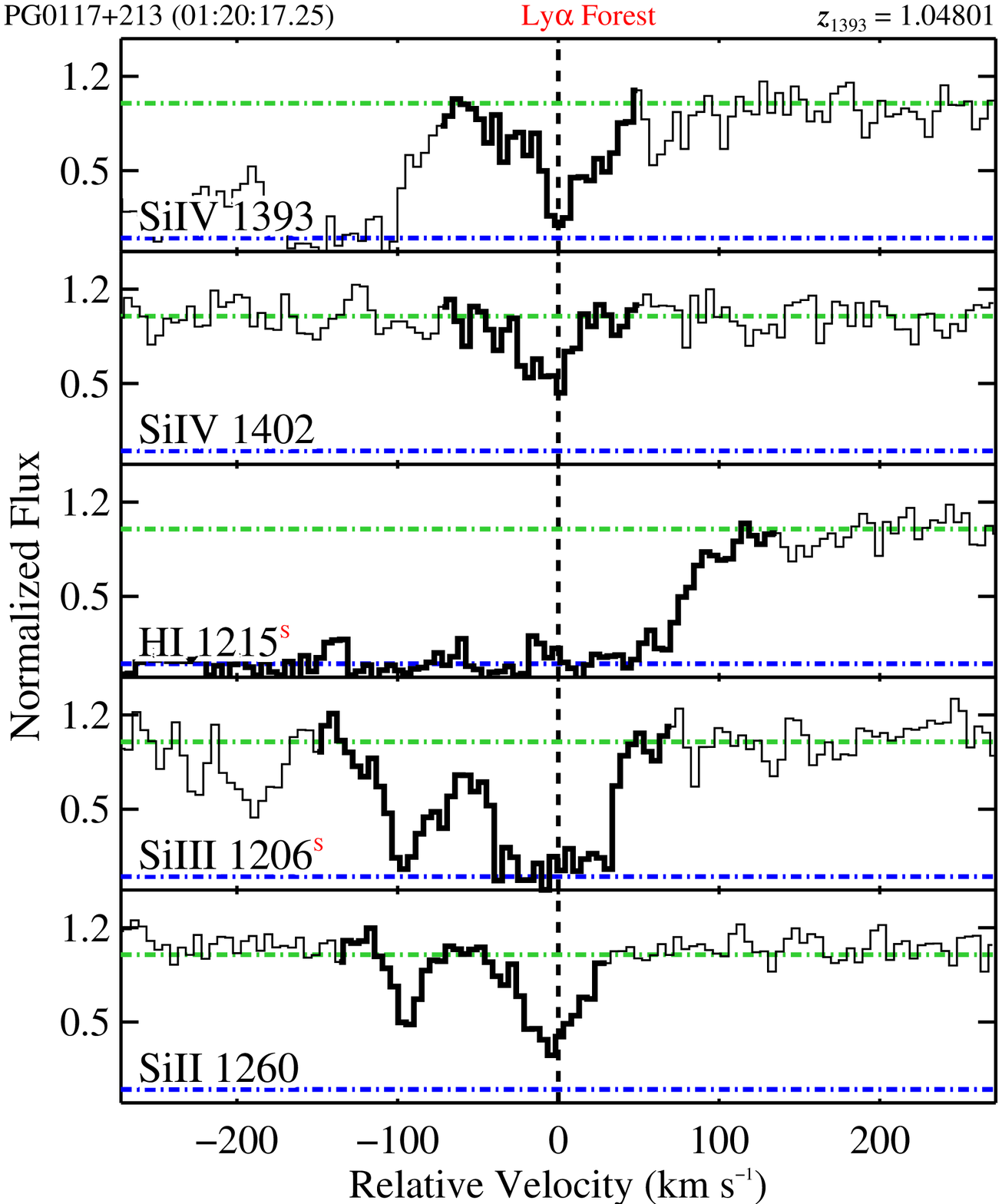} 
  \caption[Velocity plots of 22 G=1 \ion{Si}{4} systems.]
  {Velocity plots of 22 G = 1 \ion{Si}{4} systems. The regions of
    spectra around each absorption line are aligned in velocity space
    with respect to the rest wavelength of the transition and
    \zsiiv. Saturated transitions are indicated with the (red) `S';
    any transition with column density detected at $< 3\,\sigma$ are
    noted with a (red) `W.' Systems with $\zsiiv <
    \lambda_{\alpha}(1+\zem)$ are labeled with a (red) ``\Lya\
    Forest.'' The regions used to measure \EWr\ and \logSiIV\ are
    shown by the dark outline. The flux at zero and unity are shown
    with the dash-dot lines (blue and green, respectively); the
    vertical dashed line indicates $v=0\kms$, corresponding to the
    optical depth-weighted velocity centroid of the \ion{Si}{4} 1393
    transition. (The G = 1 velocity plots are available in their
    entirety in the online journal. One figure is shown here for
    guidance regarding the form and content.)
    \label{fig.g1}
  }
\end{figure}

All candidates were visually inspected by at least one author, and the
candidates with both lines detected at $\EWr \ge 3\sigEWr$ were
reviewed by two or more. We agreed upon 22 definite \ion{Si}{4}
systems, which constitute the ``G = 1'' group (see Figure
\ref{fig.g1}). Of these, 20 have both doublet lines detected with rest
equivalent width $\EWr \ge 3\sigEWr$ and constitute the group on
which we based our conclusions. The G = 1 sample has a median redshift
$\langle z \rangle = 0.906$, $\logSiIV > 12.9$, and $\EWlin{1393} \ge
66\mA$.

\begin{figure}[!hbt]
  \includegraphics[width=0.45\textwidth]{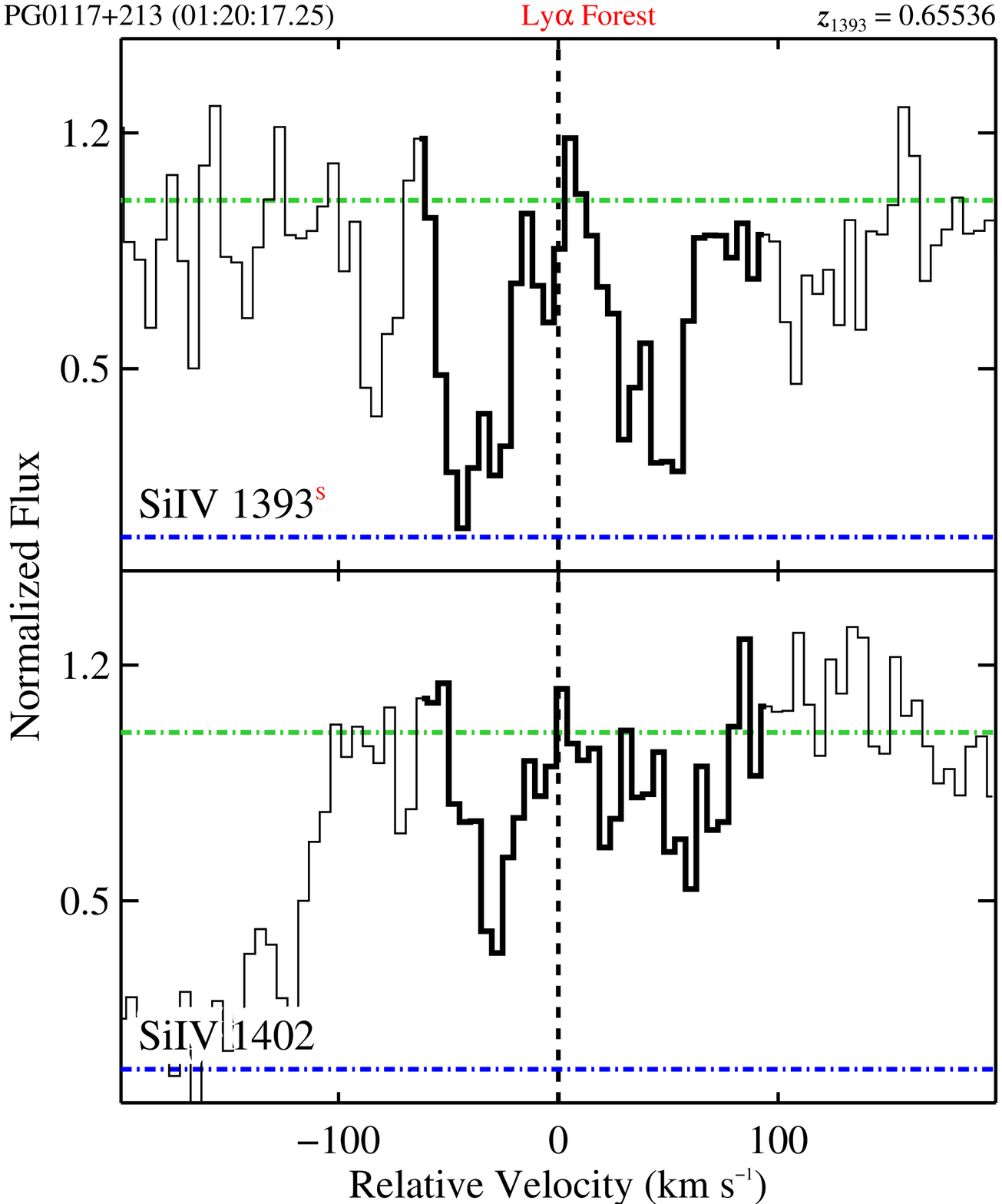} 
  \caption[Velocity plots of six G = 2 \ion{Si}{4} systems.]
  {Velocity plots of six G = 2 \ion{Si}{4} systems. (See Figure
    \ref{fig.g1} for description of velocity plot.)  (The G = 2
    velocity plots are available in their entirety in the online
    journal. One figure is shown here for guidance regarding the form
    and content.)
    \label{fig.g2}
  }
\end{figure}

We also defined a small, ``highly-likely'' (G = 2) sample of six
systems (see Figure \ref{fig.g2}). These doublets are typically found
in regions with low S/N and\slash or do not have other lines
associated with them, which would increase the confidence of our
identification. The four of these with both lines detected at $\ge
3\sigEWr$ were combined with the G = 1 sample for some analyses and
identified in tables and figures as G = 1+2. Details of all (G = 1+2)
absorption systems are given in Table \ref{tab.sys}. The properties of
the all \ion{Si}{4} doublets are summarized in Table \ref{tab.siiv}.

All doublets are more than $1000\kms$ outside of the Galaxy and
$3000\kms$ blue-ward of the background quasar.  There were no
\ion{Si}{4} absorbers without associated \Lya\ absorption, when the
spectral coverage existed to detect \Lya.  We combined \ion{Si}{4}
doublets into one absorption system when their optical depth-weighted
centroids had $\dvabs < 250\kms$.

The observed absorber line density is the sum of the number of
absorbers, each weighted by the path length sensitive to their \NSiIV\
or \EWlin{1393}. For the G = 1 sample, $\dNSiIVdX = 1.4^{+0.4}_{-0.3}$
($\dNSiIVdz = 2.7^{+0.7}_{-0.6}$) for $\logSiIV > 12.9$. 

As in \citetalias{cookseyetal10}, we conducted Monte Carlo simulations
to measure the rate that pairs of $z \lesssim 1.5$ \ion{H}{1} Lyman
forest lines satisfy the characteristics of \ion{Si}{4} doublets and
were potentially included in our sample as doublets. The contamination
rate of forest lines masquerading as \ion{Si}{4} was small, less than
5\% of \dNSiIVdX\ or an expected $1^{+2}_{-1}$ false doublet. If any
doublet in our sample were false, it would be one without other
associated absorption lines. There are four such doublets, all in the
G = 2 sample. Our expectation that forest lines could mimic \ion{Si}{4}
doublets drove us to define the ``highly-likely'' G = 2 group. Though
we provide the results for analyses of the G = 1+2 sample, we only
discuss the results from the G = 1 sample and base our conclusions on
that; so we concern ourselves no further with the effects of the
\ion{H}{1} Lyman forest contamination.

\subsection{Comparison with Previous
  Studies}\label{subsec.compstudies} 

There have been three recent surveys for \ion{Si}{4} systems at $z
\lesssim 1$ using at least some of the \hst\ spectra analyzed here:
\citet{milutinovicetal07, danforthandshull08}, and
\citetalias{cookseyetal10}. Here we briefly compare our blind doublet
search results with these other studies. In the first two, they
identified \Lya\ lines first and then sought associated transitions

\citet{milutinovicetal07} identified 17 \ion{Si}{4} doublets in the
eight \stis\ E230M spectra that they surveyed. We independently
identified 13 of those absorbers.  The remaining four doublets were
explicitly listed as questionable by these authors. We identified the
G = 2, $\zsiiv = 0.65536$ absorber in the PG0117+213 spectrum, one of
the eight surveyed by \citet{milutinovicetal07}, though they did not
detect it. Since only the doublet was detected, they would not have
found it with their \Lya-targeted search.

\citet{danforthandshull08} surveyed all of the \stis\ E140M spectra
for many transitions, including \ion{Si}{4}. They did not require that
both doublet lines be detected with $\EWr \ge 3\sigEWr$. They found 20
\ion{Si}{4} doublets for a line density $\dNSiIVdX = 5^{+2}_{-1}$ for
$\EWr \ge 30\mA$.

In contrast, we found seven \ion{Si}{4} absorbers in the E140M
spectra.  There were two (G = 1) doublets that we identified in our
\ion{Si}{4}-targeted survey but \citet{danforthandshull08} did not
find: the $\zsiiv = 0.00572$ doublet in QSO--123050+011522 and the
$\zsiiv = 0.13846$ one in PG1116+215.  As mentioned in
\citetalias{cookseyetal10}, \citet{danforthandshull08} missed the
former doublet because the \Lyb\ line in the \fuse\ spectra was
suspect. The latter doublet has $\EWlin{1402} < 3\sigEWlin{1402}$,
which might explain why they did not identify it.

Of their 20 systems, we agree with five and independently identified
them in our \ion{Si}{4}-targeted survey.  The remaining 15 doublets
from \citet{danforthandshull08} were not included in our sample for at
least one of the following reasons: one or both lines detected were at
$\EWr < 3\sigEWr$ (12 doublets); one line was blended with a Galactic
line (1); it was a system intrinsic to the background quasar (1);
and\slash or the doublet was observed in the 0th order of E140M (2),
which was exclude in our reduction.

We recovered all \ion{Si}{4} doublets that we previously identified
in \citetalias{cookseyetal10} because of the association with a
\ion{C}{4} system. We also determined that the G = 1, $z_{1548} =
0.24010$ \ion{C}{4} doublet from \citetalias{cookseyetal10} is
actually \ion{O}{1} $\lambda 1302$ and \ion{Si}{2} $\lambda 1304$
associated with the G = 1, $z_{1548} = 0.47436$ system, which
corresponds to our G = 1 \ion{Si}{4} system with $\zsiiv = z_{1548}$.



\begin{deluxetable}{rllllllllll}
\tablewidth{0pc}
\tablecaption{\ion{Si}{4} SYSTEMS SUMMARY \label{tab.sys}}
\tabletypesize{\scriptsize}
\tablehead{ 
\colhead{(1)} & \colhead{(2)} & \colhead{(3)} & \colhead{(4)} & \colhead{(5)} & 
\colhead{(6)} & \colhead{(7)} & \colhead{(8)} & \colhead{(9)} & \colhead{(10)} & 
\colhead{(11)} \\
\colhead{$z_{1393}$} & \colhead{\dvabs} & 
\colhead{$\lambda_{r}$} & \colhead{\wvlo} & 
\colhead{\wvhi} & \colhead{\EWr} & \colhead{\sigEWr} & 
\colhead{\logN} & \colhead{\siglogN} & 
\colhead{G} & \colhead{Flag} \\ 
 & \colhead{(\!\kms)} & \colhead{(\AA)} & \colhead{(\AA)} & 
\colhead{(\AA)} & \colhead{(\!\mA)} & \colhead{(\!\mA)} & & & & 
}
\startdata
\hline \\[-1ex]
\multicolumn{11}{c}{PG0117+213 ($\zem=1.493$)} \\[1ex]
\hline
0.65536 & \nodata & 1393.76 & 2306.70 & 2307.89 &   273 &    28 & $> 13.73$ & \nodata &  2 & 482 \\
 & $    9.2$ & 1402.77 & 2321.62 & 2322.82 &   122 &    24 & 13.54 &  0.09 &      &   \\
1.04801 & \nodata & 1393.76 & 2853.75 & 2854.89 &   194 &    18 & 13.54 &  0.07 &  1 & 471 \\
 & $  -10.5$ & 1402.77 & 2872.21 & 2873.36 &    83 &    14 & 13.37 &  0.07 &      &   \\
 & $ -111.3$ & 1215.67 & 2486.96 & 2490.84 &  1530 &    21 & $> 14.88$ & \nodata &      &   \\
 & $  -27.2$ & 1206.50 & 2469.68 & 2471.49 &   446 &    19 & $> 13.67$ & \nodata &      &   \\
 & $  -26.2$ & 1260.42 & 2580.19 & 2581.61 &   145 &    11 & 13.16 &  0.03 &      &   \\
\hline \\[-1ex]
\multicolumn{11}{c}{B0312--770 ($\zem=0.223$)} \\[1ex]
\hline
0.20253 & \nodata & 1393.76 & 1674.90 & 1676.82 &   652 &    78 & $> 14.16$ & \nodata &  1 & 511 \\
 & $    7.6$ & 1402.77 & 1685.73 & 1687.67 &   543 &    57 & $> 14.40$ & \nodata &      &   \\
 & $  -15.0$ & 1215.67 & 1460.66 & 1463.09 &  1677 &    20 & $> 14.92$ & \nodata &      &   \\
 & $   -7.3$ & 1025.72 & 1232.52 & 1234.29 &  1139 &    20 & $> 15.59$ & \nodata &      &   \\
 & $   11.2$ &  977.02 & 1174.06 & 1175.78 &   912 &    67 & $> 14.44$ & \nodata &      &   \\
 & $   51.6$ & 1031.93 & 1240.64 & 1241.62 &   496 &    17 & $> 14.94$ & \nodata &      &   \\
 & $   58.2$ & 1037.62 & 1247.48 & 1248.47 &   310 &    14 & 14.87 &  0.02 &      &   \\
 & $   21.7$ & 1206.50 & 1449.82 & 1451.60 &   763 &    19 & $> 13.94$ & \nodata &      &   \\
 & $   61.5$ & 1238.82 & 1489.57 & 1490.45 &   137 &     7 & 13.86 &  0.02 &      &   \\
 & $   62.1$ & 1242.80 & 1494.36 & 1495.24 &    94 &     9 & 13.99 &  0.04 &      &   \\
 & $   39.1$ & 1260.42 & 1515.44 & 1516.45 &   490 &    12 & $> 13.92$ & \nodata &      &   \\
\hline \\[-1ex]
\multicolumn{11}{c}{PKS0405--123 ($\zem=0.573$)} \\[1ex]
\hline
0.16712 & \nodata & 1393.76 & 1626.40 & 1626.95 &   117 &    15 & 13.24 &  0.06 &  1 & 499 \\
 & $    5.4$ & 1402.77 & 1636.92 & 1637.47 &   103 &    23 & 13.49 &  0.10 &      &   \\
 & $  -18.1$ & 1215.67 & 1417.80 & 1419.38 &   876 &    19 & $> 14.60$ & \nodata &      &   \\
 & $   -0.9$ & 1025.72 & 1196.77 & 1197.50 &   444 &    20 & $> 15.19$ & \nodata &      &   \\
 & $  -22.2$ &  977.02 & 1139.53 & 1140.73 &   500 &    22 & $> 14.26$ & \nodata &      &   \\
 & $   12.1$ &  989.80 & 1154.85 & 1155.66 &   214 &    12 & $> 14.64$ & \nodata &      &   \\
 & $  -30.1$ & 1031.93 & 1203.74 & 1204.67 &   391 &    24 & $> 14.77$ & \nodata &      &   \\
 & $  -50.7$ & 1037.62 & 1210.37 & 1211.31 &   218 &    37 & 14.80 &  0.11 &      &   \\
 & $    2.3$ & 1206.50 & 1407.81 & 1408.40 &   236 &    11 & $> 13.38$ & \nodata &      &   \\
 & $   -6.4$ & 1238.82 & 1445.53 & 1446.12 &   108 &    15 & 13.81 &  0.06 &      &   \\
 & $   -7.4$ & 1242.80 & 1450.17 & 1450.76 &    76 &    13 & 13.94 &  0.07 &      &   \\
 & $   -3.3$ & 1260.42 & 1470.73 & 1471.28 &   148 &    12 & $> 13.28$ & \nodata &      &   \\
\enddata
\tablecomments{Si\,IV systems by target and redshift of Si\,IV 1393.
Upper limits are $2\sigma$ limits for both \EWr\ and \logN.
The column densities were measured by the AODM, unless ``COG'' is indicated, in which case, the limit is from assuming \EWr\ results from the linear portion of the COG.
The definite Si\,IV doublets are labeled group G = 1, while the ``highly-likely'' doublets are G = 2.
The binary flag is described in Section \ref{sec.selec}.
\\
(This table is available in a machine-readable form in the online journal.  A portion is shown here for guidance regarding its form and content.)
}
\end{deluxetable}



\begin{deluxetable}{llrr@{}lr@{}lr@{}lr@{}lr@{}l}
\tablecolumns{13}
\tablewidth{0pc}
\tablecaption{\ion{Si}{4} DOUBLET SUMMARY \label{tab.siiv}}
\tabletypesize{\scriptsize}
\tablehead{ 
\colhead{(1)} & \colhead{(2)} & \colhead{(3)} & 
\multicolumn{2}{c}{(4)} & \multicolumn{2}{c}{(5)} & 
\multicolumn{2}{c}{(6)} & \multicolumn{2}{c}{(7)} & 
\multicolumn{2}{c}{(8)} \\
\colhead{Target} & \colhead{G} & \colhead{$z_{1393}$} & 
\multicolumn{2}{c}{$\EWlin{1393}$} & \multicolumn{2}{c}{$\EWlin{1402}$} & 
\multicolumn{2}{c}{$\logN{}_{1393}$} & \multicolumn{2}{c}{$\logN{}_{1402}$} & 
\multicolumn{2}{c}{$\logSiIV$} \\ 
 & & & \multicolumn{2}{c}{(\mA)} & \multicolumn{2}{c}{(\mA)} & 
\multicolumn{2}{c}{} & \multicolumn{2}{c}{} & \multicolumn{2}{c}{} 
}
\startdata
                    PG0117+213 & 2 & 0.65536 & $  273$&$\,\pm\,   28$ & $  122$&$\,\pm\,   24$ & $> 13.73$ & & $13.54$&$\,\pm\, 0.09$ & $13.54$&$\,\pm\, 0.09$ \\
 & 1 & 1.04801 & $  194$&$\,\pm\,   18$ & $   83$&$\,\pm\,   14$ & $13.54$&$\,\pm\, 0.07$ & $13.37$&$\,\pm\, 0.07$ & $13.43$&$\,\pm\, 0.05$ \\
                    B0312--770 & 1 & 0.20253 & $  652$&$\,\pm\,   78$ & $  543$&$\,\pm\,   57$ & $> 14.16$ & & $> 14.40$ & & $> 14.40$ & \\
                  PKS0405--123 & 1 & 0.16712 & $  117$&$\,\pm\,   15$ & $  103$&$\,\pm\,   23$ & $13.24$&$\,\pm\, 0.06$ & $13.49$&$\,\pm\, 0.10$ & $13.27$&$\,\pm\, 0.05$ \\
                   PKS0454--22 & 1 & 0.47436 & $  635$&$\,\pm\,   52$ & $  475$&$\,\pm\,   56$ & $> 13.84$\tablenotemark{a} & & $> 14.02$\tablenotemark{a} & & $> 14.02$ & \\
 & 1 & 0.48331 & $  112$&$\,\pm\,   37$ & $<   61$ & & $> 13.09$\tablenotemark{a} & & $< 13.34$ & & $[13.09$&$,13.38]$ \\
                  HE0515--4414 & 2 & 0.68625 & $  249$&$\,\pm\,   18$ & $   68$&$\,\pm\,   17$ & $< 13.71$\tablenotemark{b} & & $13.36$&$\,\pm\, 0.10$ & $13.36$&$\,\pm\, 0.10$ \\
 & 1 & 0.94047 & $  233$&$\,\pm\,   13$ & $   56$&$\,\pm\,   11$ & $< 13.58$\tablenotemark{b} & & $13.15$&$\,\pm\, 0.08$ & $13.15$&$\,\pm\, 0.08$ \\
 & 1 & 1.14760 & $  565$&$\,\pm\,   22$ & $  343$&$\,\pm\,   29$ & $13.98$&$\,\pm\, 0.03$ & $13.98$&$\,\pm\, 0.04$ & $13.98$&$\,\pm\, 0.02$ \\
                   HS0810+2554 & 2 & 0.55455 & $  250$&$\,\pm\,   54$ & $<  160$ & & $> 13.44$\tablenotemark{a} & & $< 14.06$ & & $[13.44$&$,13.76]$ \\
 & 2 & 0.83143 & $  262$&$\,\pm\,   30$ & $  271$&$\,\pm\,   31$ & $13.67$&$\,\pm\, 0.08$ & $13.91$&$\,\pm\, 0.06$ & $13.77$&$\,\pm\, 0.05$ \\
                       MARK132 & 2 & 0.93166 & $   73$&$\,\pm\,   12$ & $   86$&$\,\pm\,   12$ & $12.95$&$\,\pm\, 0.07$ & $< 13.31$\tablenotemark{b} & & $12.95$&$\,\pm\, 0.07$ \\
                    PG1116+215 & 1 & 0.13846 & $   34$&$\,\pm\,    9$ & $<   18$ & & $12.68$&$\,\pm\, 0.10$ & $< 12.64$ & & $12.68$&$\,\pm\, 0.10$ \\
                    PG1206+459 & 1 & 0.92690 & $  777$&$\,\pm\,   19$ & $  402$&$\,\pm\,   29$ & $14.08$&$\,\pm\, 0.01$ & $14.06$&$\,\pm\, 0.03$ & $14.08$&$\,\pm\, 0.01$ \\
 & 1 & 0.93429 & $  143$&$\,\pm\,   11$ & $   98$&$\,\pm\,   11$ & $> 13.51$ & & $13.60$&$\,\pm\, 0.06$ & $13.60$&$\,\pm\, 0.06$ \\
                    PG1211+143 & 1 & 0.05118 & $   66$&$\,\pm\,    4$ & $   44$&$\,\pm\,    5$ & $12.96$&$\,\pm\, 0.03$ & $13.04$&$\,\pm\, 0.05$ & $12.97$&$\,\pm\, 0.03$ \\
                        MRK205 & 1 & 0.00428 & $  122$&$\,\pm\,   11$ & $   75$&$\,\pm\,   13$ & $13.25$&$\,\pm\, 0.04$ & $13.33$&$\,\pm\, 0.07$ & $13.27$&$\,\pm\, 0.04$ \\
            QSO--123050+011522 & 1 & 0.00572 & $   66$&$\,\pm\,    8$ & $   63$&$\,\pm\,    9$ & $12.96$&$\,\pm\, 0.05$ & $13.21$&$\,\pm\, 0.06$ & $13.02$&$\,\pm\, 0.04$ \\
                    PG1248+401 & 1 & 0.77305 & $  472$&$\,\pm\,   19$ & $  348$&$\,\pm\,   17$ & $> 14.01$ & & $14.06$&$\,\pm\, 0.02$ & $14.06$&$\,\pm\, 0.02$ \\
 & 1 & 0.85485 & $  211$&$\,\pm\,   21$ & $  138$&$\,\pm\,   31$ & $13.49$&$\,\pm\, 0.05$ & $13.59$&$\,\pm\, 0.10$ & $13.50$&$\,\pm\, 0.04$ \\
                    PG1630+377 & 1 & 0.91432 & $  123$&$\,\pm\,   15$ & $   98$&$\,\pm\,   12$ & $13.26$&$\,\pm\, 0.06$ & $13.47$&$\,\pm\, 0.06$ & $13.33$&$\,\pm\, 0.04$ \\
 & 1 & 0.95279 & $  202$&$\,\pm\,   15$ & $  136$&$\,\pm\,   21$ & $13.49$&$\,\pm\, 0.04$ & $13.59$&$\,\pm\, 0.07$ & $13.51$&$\,\pm\, 0.03$ \\
                    PG1634+706 & 1 & 0.65351 & $  145$&$\,\pm\,   11$ & $   71$&$\,\pm\,    9$ & $13.30$&$\,\pm\, 0.04$ & $13.25$&$\,\pm\, 0.06$ & $13.29$&$\,\pm\, 0.03$ \\
 & 2 & 0.81813 & $<   15$ & & $   21$&$\,\pm\,    7$ & $< 12.25$ & & $12.71$&$\,\pm\, 0.13$ & $12.71$&$\,\pm\, 0.13$ \\
 & 1 & 0.90560 & $  182$&$\,\pm\,    4$ & $   32$&$\,\pm\,    5$ & $< 13.68$\tablenotemark{b} & & $12.92$&$\,\pm\, 0.07$ & $12.92$&$\,\pm\, 0.07$ \\
 & 1 & 0.99035 & $  240$&$\,\pm\,    7$ & $  162$&$\,\pm\,    6$ & $13.73$&$\,\pm\, 0.02$ & $13.74$&$\,\pm\, 0.02$ & $13.74$&$\,\pm\, 0.01$ \\
 & 1 & 1.04106 & $  167$&$\,\pm\,    5$ & $   96$&$\,\pm\,    8$ & $13.32$&$\,\pm\, 0.02$ & $13.35$&$\,\pm\, 0.03$ & $13.33$&$\,\pm\, 0.01$ \\
                       PHL1811 & 1 & 0.08094 & $  108$&$\,\pm\,    7$ & $   74$&$\,\pm\,    9$ & $> 13.44$ & & $13.45$&$\,\pm\, 0.05$ & $13.45$&$\,\pm\, 0.05$ \\
\enddata
\tablenotetext{a}{\logN\ measured by assuming \EWr\ results from the linear portion of the COG.}
\tablenotetext{b}{\!\,Limit due to blended line.}
\tablecomments{
Summary of Si\,IV doublets by target and redshift of Si\,IV 1393.
The definite Si\,IV doublets are labeled group G = 1, while the ``highly-likely'' doublets are G = 2.
Upper limits are $2\sigma$ limits for both \EWr\ and \logN.
The adopted column density for the Si\,IV doublets are listed in the last column (see Section \ref{sec.data}).
}
\end{deluxetable}

\section{Analysis}\label{sec.analysis}

\subsection{Frequency Distributions}\label{subsec.freqdistr}

\begin{figure}[!hbt]
  \begin{center}$
    \begin{array}{c}
     \includegraphics[height=0.45\textwidth,angle=90]{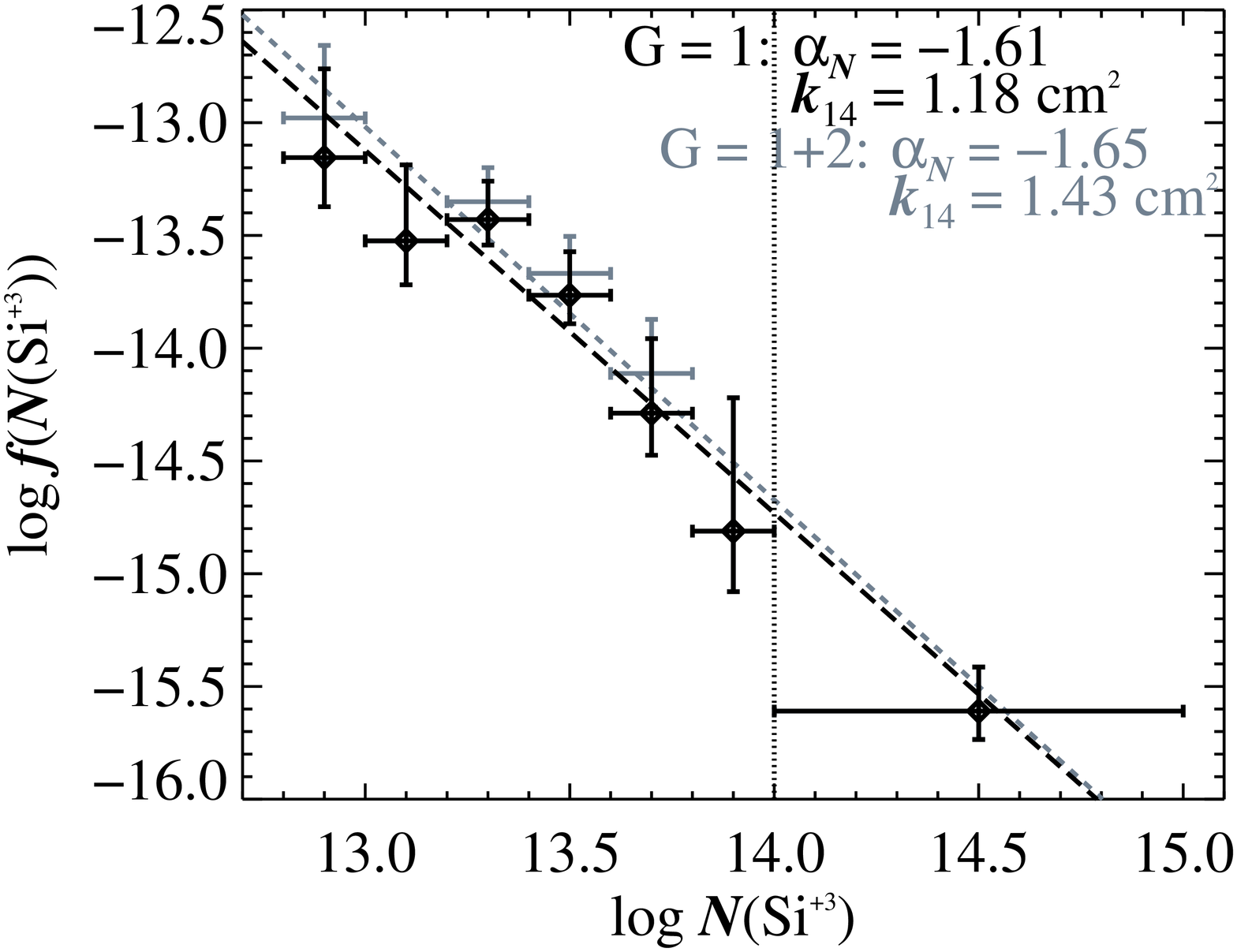} \\ 
     \includegraphics[height=0.45\textwidth,angle=90]{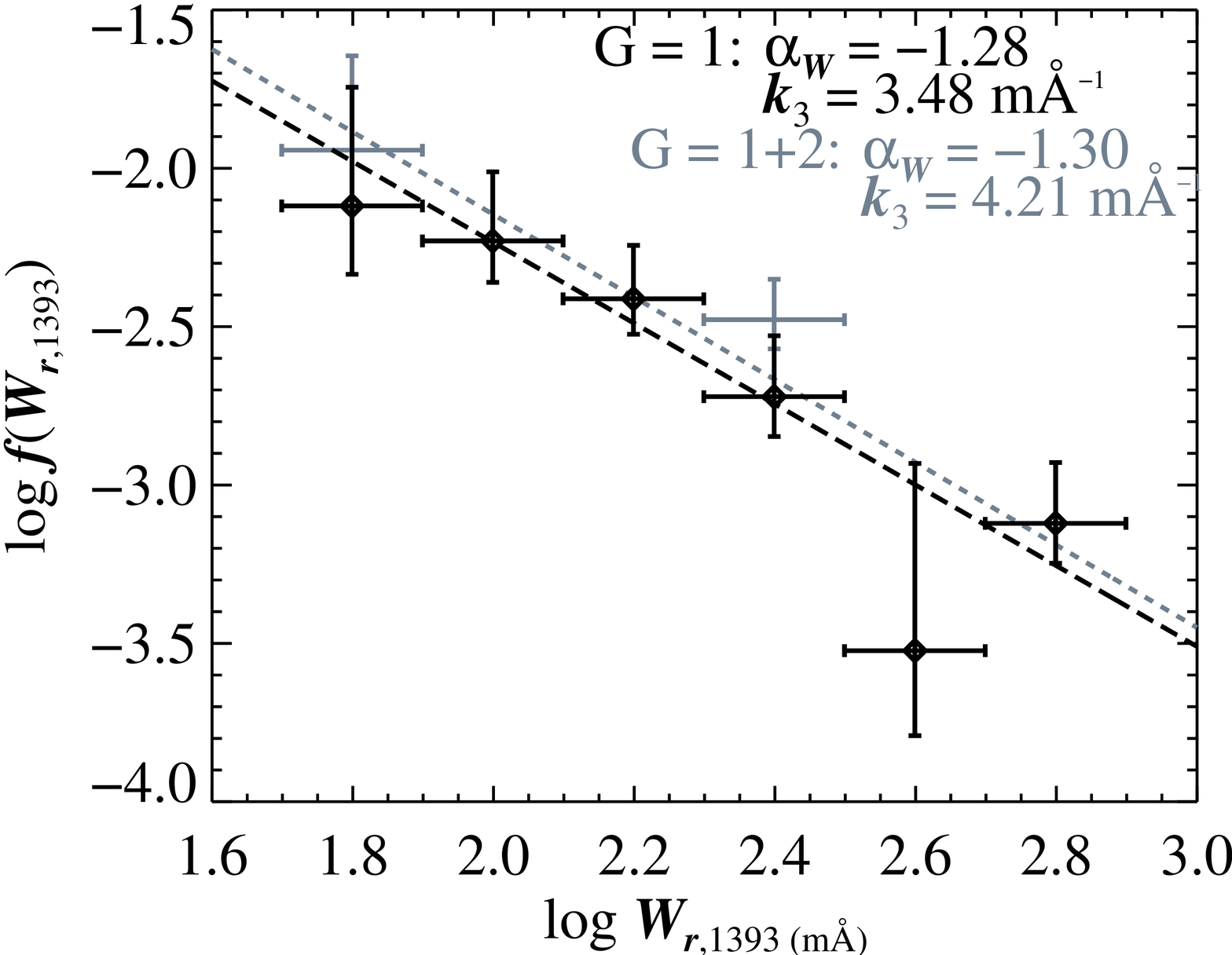} 
   \end{array}$
\end{center}
  \caption[Column density (top) and rest equivalent width (bottom)
  frequency distributions for G = 1 and 1+2 samples.] 
  {Column density (top) and rest equivalent width (bottom) frequency
    distributions for G = 1 and 1+2 samples (black and gray,
    respectively). The best-fit power laws, from the maximum
    likelihood analysis, are the dashed lines, with slope $\alpha$ and
    normalization \kff{14}\ as given.  The column density saturation
    limit is indicated by the vertical, dotted line (top).
    \label{fig.fn} 
  }
\end{figure}

Analogous to the luminosity function used in galaxy studies, observers
define the column density frequency distribution \ff{\NSiIV} to be the
number $\Delta \Num$ of \ion{Si}{4} doublets per column density
interval $\Delta \NSiIV$ per path length \DXp\ (see Figure
\ref{fig.fn}). With a maximum likelihood fitting algorithm, we fit a
power law to \ff{\NSiIV} as follows:
\begin{equation}
  \ff{\NSiIV} = k\,\bigg(\frac{\NSiIV}{\N{0}}\bigg)^{\aff{N}}{\rm ,} \label{eqn.fnpl} 
\end{equation}
where $k$ is the normalization with unit \!\cm{2};
$\N{0}=10^{13.5}\cm{-2}$; and \aff{N} is the slope. We often refer to
\kff{14}, which is the normalization $k$ scaled up by a factor of
$10^{14}$. The frequency distribution was fit over the range $\log
\N{\rm min} = 12.84$ to $\log \N{\rm max} = 15$, with special
treatment of the saturated absorbers ($\logN{_{\rm sat}} = 14$) in the
maximum likelihood analysis
\citepalias[see][]{cookseyetal10}. Briefly, the number of doublets
with $14 \le \logSiIV \le 15$ (\ie saturated) was a constraint in our
likelihood function. The best-fit parameters were: $\kff{14} =
1.18^{+0.45}_{-0.36}\cm{2}$ and $\aff{N} = -1.61^{+0.28}_{-0.31}$.

The fit parameters for \ff{\NSiIV} and \ff{\EWlin{1393}} (discussed
below) are given in Table \ref{tab.freqdistr}.  We estimated the
68.3\% confidence limits (c.l.) in the power-law normalization and
slope by tracing a contour where $\delta \mathcal{L} \equiv \ln
\mathcal{L} - \ln \mathcal{L}_{\rm max} = -1.15$ on the likelihood
surface $\mathcal{L}$, which included $68.3\%$ of its area. Then, the
68.3\% c.l. (what we will loosely refer to as ``1-$\sigma$ errors,''
hereafter) were defined as the difference between the $k$ and $\alpha$
extrema on the contour and the $\mathcal{L}_{\rm max}$ values. Errors
in quantities derived from the frequency distributions (\eg the
\Sithr\ mass density discussed below) are estimated in a similar
fashion.  The likelihood surface is not Gaussian, and the ``2- and
3-$\sigma$ errors'' are defined by $\delta \mathcal{L} = -3.15$
(95.4\% c.l.) and $-6.1$ (99.7\% c.l.), respectively.  Therefore, the
larger confidence limits on the best-fit power-law parameters are,
formally, as follows: $2\sigma_{\kff{14}} = +0.79/-0.54\cm{2}$;
$2\sigma_{\aff{N}} = +0.45/-0.53$; $3\sigma_{\kff{14}} =
+1.18/-0.70\cm{2}$; and $3\sigma_{\aff{N}} = +0.62/-0.76$.

There was no observed break in \ff{\NSiIV}, and no break has been
observed at high redshift. There must be a break in order to limit the
number and mass of \ion{Si}{4} absorbers to finite quantities.

We have measured a slope consistent with those from high-redshift
studies, though potentially shallower.  \citet{songaila97} measured
$\aff{N} = -1.8$ for $2.16 \le z \le 3$ and $-2$ for $3 \le z \le
3.54$. \citet{songaila01} and \citet{scannapiecoetal06} stated that
$\aff{N}=-1.8$ matched their observed $\ff{\NSiIV}$ well, which
covered $1.78 \le z \le 5.29$ and $1.5 \le z \le 3.1$, respectively.
From their \Lya-targeted $z < 0.4$ survey, \citet{danforthandshull08}
measured $\aff{N} = -1.92\pm0.17$ for a sample with different
selection criteria and from a survey with different methodology. 

The definition of the equivalent width frequency distribution
\ff{\EWlin{1393}} is similar to that of \ff{\NSiIV}, and it was also
fit well with a power law, with $\EWlin{0} = 150\mA$, $\EWlin{min} =
57\mA$, and $\EWlin{max} = 796\mA$. The \EWr\ limits reflect the
extrema of the observed values, $\pm1\sigma$. The best-fit values for
\ff{\EWlin{1393}} were: $\kff{3} = 3.48^{+1.41}_{-1.12}\mA^{-1}$ and
$\aff{W} = -1.28^{+0.47}_{-0.47}$. The \EWr\ \kff{3} is the
normalization scaled up by a factor of $10^{3}$. The larger confidence
limits on the best-fit power-law parameters for \ff{\EWlin{1393}} are,
formally, as follows: $2\sigma_{\kff{3}} = +2.55/\!-\!1.73\mA^{-1}$;
$2\sigma_{\aff{W}} = +0.79/\!-\!0.80$; $3\sigma_{\kff{3}} =
+3.81/\!-\!2.21\mA^{-1}$; and $3\sigma_{\aff{W}} = +1.10/\!-\!1.13$.


\begin{deluxetable}{cccccccccccc}
\rotate
\tablewidth{0pc}
\tablecaption{\ion{Si}{4} FREQUENCY DISTRIBUTIONS SUMMARY \label{tab.freqdistr}}
\tabletypesize{\scriptsize}
\tablehead{ 
\colhead{(1)} & \colhead{(2)} & \colhead{(3)} & \colhead{(4)} & \colhead{(5)} & 
\colhead{(6)} & \colhead{(7)} & \colhead{(8)} & \colhead{(9)} & \colhead{(10)} & 
\colhead{(11)} & \colhead{(12)} \\
\colhead{G} & \colhead{$\langle z \rangle$} & \colhead{$z_{l}$} & \colhead{$z_{h}$} & 
\colhead{\Num} & \colhead{Limits} & 
\colhead{\dNSiIVdz} & \colhead{\dNSiIVdX} & 
\colhead{$\OmSiIV\times10^{8}$} & 
\colhead{$k$\tablenotemark{a}} & 
\colhead{$\alpha$} & \colhead{$P_{\rm KS}$} 
}
\startdata
\hline \\[-1ex]
\multicolumn{12}{c}{Column Density} \\[1ex]
\hline
1 &   0.90560 &   0.00428 &   1.14760 & 20 & (12.84,\,15.00) & $ 2.7^{+ 1.1}_{- 0.8}$ & $ 1.2^{+ 0.5}_{- 0.4}$ & $ 3.71^{+ 2.82}_{- 1.68}$ & $ 1.18^{+ 0.45}_{- 0.36}$ & $-1.61^{+ 0.28}_{- 0.31}$ & 0.526 \\
 &   0.91432 & & & 18 & (12.92,\,14.40) & $ 2.4^{+ 0.4}_{- 0.3}$ & $ 1.4^{+ 0.2}_{- 0.2}$ & $> 1.63$ & \nodata &  \nodata &  \nodata \\
1+2 &   0.85485 &   0.00428 &   1.14760 & 24 & (12.84,\,15.00) & $ 3.1^{+ 1.1}_{- 0.9}$ & $ 1.5^{+ 0.5}_{- 0.4}$ & $ 4.13^{+ 2.81}_{- 1.77}$ & $ 1.43^{+ 0.48}_{- 0.39}$ & $-1.65^{+ 0.26}_{- 0.28}$ & 0.400 \\
 &   0.90560 & & & 22 & (12.92,\,14.40) & $ 2.9^{+ 0.4}_{- 0.3}$ & $ 1.7^{+ 0.2}_{- 0.2}$ & $> 1.94$ & \nodata &  \nodata &  \nodata \\
\hline \\[-1ex]
\multicolumn{12}{c}{Equivalent Width} \\[1ex]
\hline
1 &   0.90560 &   0.00428 &   1.14760 & 20 & $(  57,\, 796)$ & $ 3.0^{+ 1.2}_{- 0.9}$ & $ 1.4^{+ 0.5}_{- 0.4}$ & \nodata & $ 3.48^{+ 1.41}_{- 1.12}$ & $-1.28^{+ 0.47}_{- 0.47}$ & 0.533 \\
 &  & & & & $(  66,\, 777)$ & $ 2.2^{+ 0.3}_{- 0.3}$ & $ 1.3^{+ 0.2}_{- 0.1}$ & \nodata &  \nodata &  \nodata &  \nodata \\
1+2 &   0.85485 &   0.00428 &   1.14760 & 24 & $(  57,\, 796)$ & $ 3.5^{+ 1.3}_{- 1.0}$ & $ 1.6^{+ 0.6}_{- 0.5}$ & \nodata & $ 4.21^{+ 1.53}_{- 1.24}$ & $-1.30^{+ 0.43}_{- 0.44}$ & 0.563 \\
 &  & & & & $(  66,\, 777)$ & $ 2.6^{+ 0.3}_{- 0.3}$ & $ 1.5^{+ 0.2}_{- 0.2}$ & \nodata &  \nodata &  \nodata &  \nodata \\
\enddata
\tablenotetext{a}{The power-law coefficient $k$ has units of $10^{-14}\cm{2}$ for the column density section and $10^{-3}\mA^{-1}$ for the equivalent width section.}
\tablecomments{
Parameters from the maximum likelihood analysis for $\ff{x} = k\,(x/x_{0})^{\alpha}$, where $x=\NSiIV$ (or \EWlin{1393}) and $x_{0}$ is $\N{0} = 10^{13.5}\cm{-2}$ ($\EWlin{0} = 150\mA$).
For each G = 1 or 1+2 subsample, the first row summarizes the maximum likelihood analysis and the second row, the observed quantities.
\dNSiIVdX, listed in the first subsample row, is the integral of \ff{\NSiIV}\ (\ff{\EWlin{1393}}) from $\logSiIV=13$ to infinity ($\EWlin{1393}=50\mA$ to $796\mA$) with the best-fit $k$ and $\alpha$.
Also in the first subsample row, the integrated $\dNSiIVdz \equiv \dNSiIVdX \cdot \ud X/\ud z$, where the latter term is evaluated at $\langle z \rangle$.
The observed \dNSiIVdz\ and \dNSiIVdX\ are from the sum of the {\it total} number of doublets, weighted by the path length available to detect the doublet, based on its \NSiIV\ or \EWlin{1393}.
\OmSiIV, listed in the first subsample row in the column density section, is the integral of $\ff{\NSiIV}\cdot\NSiIV$ from $13 \le \logSiIV \le 15$ with the best-fit $k$ and $\alpha$.
The observed \OmSiIV\ were from the sum of the {\it unsaturated} doublets, as given by \Num.
$P_{\rm KS}$ is the significance of the one-sided Kolmogorov-Smirnov statistic of the best-fit power law.
}
\end{deluxetable}

\subsection{\ion{Si}{4} Absorber Line Density}\label{subsec.dndx} 

We measured the observed \ion{Si}{4} line density by the sum of the
number of absorbers, each weighted by the path length sensitive to
their \NSiIV\ or \EWlin{1393}\ (see Figure \ref{fig.x}). For the G = 1
sample, $\dNSiIVdX = 1.4^{+0.4}_{-0.3}$ ($\dNSiIVdz =
2.4^{+0.7}_{-0.5}$) for $\logSiIV \ge 12.9$.

\begin{figure}[!hbt]
  \begin{center}
     \includegraphics[height=0.47\textwidth,angle=90]{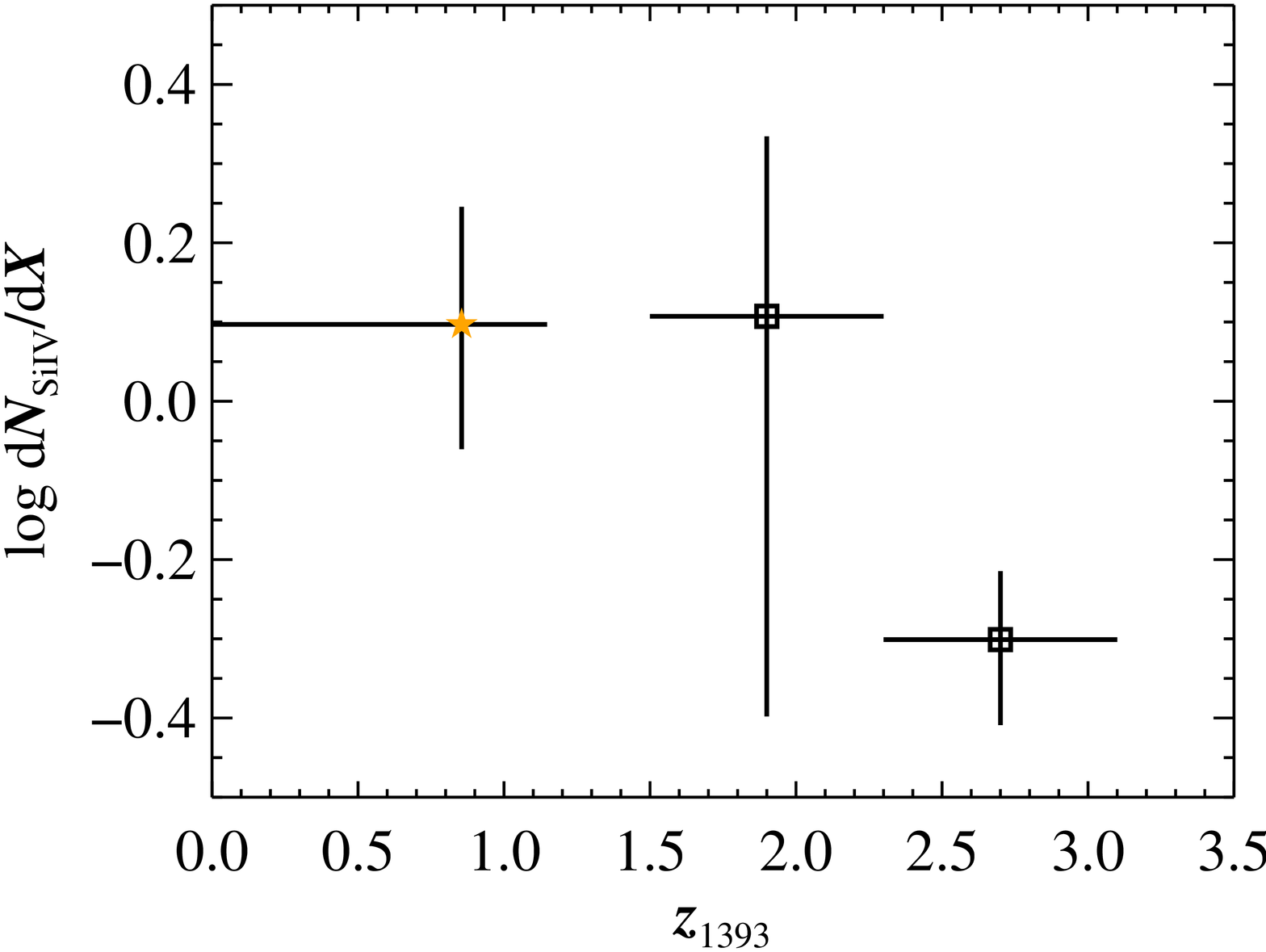} 
\end{center}
  \caption[Redshift evolution of \dNSiIVdX.]
  {Redshift evolution of \dNSiIVdX.  The line density of \ion{Si}{4}
    absorbers has not increased significantly since $z \approx 3$. The
    (orange) stars are our $z \lesssim 1$ integrated values for
    $\NSiIV \ge 10^{13}\cm{-2}$. The high-redshift values are
    integrated \dNSiIVdX, with the slope and normalization estimated
    from the \ff{\NSiIV} published in \citet[][black
    squares]{scannapiecoetal06}. 
    \label{fig.dndx} 
  }
\end{figure}

Since \ff{\NSiIV}\ is modeled well by a power law, we can integrate
Equation \ref{eqn.fnpl} to estimate the \ion{Si}{4} absorber line
density \dNSiIVdX\ for a given column density limit \N{\rm lim}:
\begin{equation} 
\frac{\displaystyle \ud \Num_{\mathrm{Si\,IV}}}{\displaystyle \ud X}
(\NSiIV \ge \N{\rm lim})  
= -\frac{k}{1 + \aff{N}}
\frac{\N{\rm lim}^{1+\aff{N}}}{\N{0}^{\aff{N}}}  {\rm .}
\label{eqn.dndx}
\end{equation}
This is useful for comparing to high-redshift studies, which typically
do not match our observational limits. In Figure \ref{fig.dndx}, we
show \dNSiIVdX\ as a function of $\zsiiv$ for $\N{\rm lim} =
10^{13}\cm{-2}$.

The \ion{Si}{4} line density does not increase significantly from $z
\approx 3 \rightarrow 0$. Our integrated, $z \lesssim 1$ $\dNSiIVdX =
1.2^{+0.5}_{-0.4}$ ($\log \N{\rm lim} = 13$) is consistent within
$1\sigma$ of the $\langle z \rangle = 1.9$ value from
\citet{scannapiecoetal06} and $2\sigma$ of their $\langle z \rangle =
2.7$ one. The formal errors on the integrated \dNSiIVdX\ are
$2\sigma_{\ud \mathcal{N}/\ud X} = +1.3/\!-\!0.6$ and $3\sigma_{\ud
  \mathcal{N}/\ud X} = +1.2/\!-\!0.8$, derived from the $\aff{N}$
and $\kff{14}$ errors discussed in Section
\ref{subsec.freqdistr}. There is no evidence of evolution in
\dNSiIVdX\ from $z = 1 \rightarrow 0$ (\ie within our sample).

We estimated the high-redshift line densities and errors in the
following manner.  \citet{scannapiecoetal06} published their
\ff{\NSiIV}, from which we estimated the power-law normalization
$\kff{14}$ given the best-fit slope $\aff{N}$. We measured
\ff{\NSiIV}\ at several \NSiIV\ and used the scatter as an estimate of
the errors.  We assumed $\aff{N} = -1.8$, as \citet{scannapiecoetal06}
did, and estimated $\kff{14} = 1.3\pm0.9\cm{2}$ for $\langle z \rangle
= 1.9$ and $\kff{14} = 0.5\pm0.1\cm{2}$ for $\langle z \rangle = 2.7$.


\subsection{\Sithr\ Mass Density}\label{subsec.omsiiv}

The \Sithr\ mass density has been measured by several $1.5 < z
\lesssim 5.5$ studies \citep{songaila97, songaila01, songaila05,
  scannapiecoetal06}. Typically, they measured the mass density,
relative to the critical density $\rho_{c,0}$, by summing the detected
absorbers \citep{lanzettaetal91}:
\begin{equation}
  \OmSiIV= \frac{H_{0}\,\mathrm{m}_{\rm C}} {c\,\rho_{c,0}}
  \sum_{\Num} \frac{\NSiIV}{\DX{\NSiIV}} {\rm ,} \label{eqn.sum_omsiiv}
\end{equation}
where the Hubble constant is $H_{0} = 70\kms\,\mathrm{Mpc}^{-1}$; the
mass of a silicon atom is $\mathrm{m}_{\rm Si} =5\times10^{-23}\,$g;
$c$ is the speed of light; and $\rho_{c,0}=9.26\times10^{-30}\,{\rm
  g}\cm{-3}$. The summed \OmSiIV\ approximates the mass density in all
absorbers with column densities within the observed \NSiIV\ range,
which is $12 \lesssim \logSiIV \lesssim 15$ for the high-redshift
studies.

Since we measured column densities with the AODM, we only have lower
limits on \NSiIV\ for the strong, saturated doublets, which dominate
\OmSiIV. Thus, summing the column densities only resulted in a lower
limit $\OmSiIV \ge 1.55\times10^{-8}$ for the 16 unsaturated G = 1
doublets with $\logSiIV \ge 13$. We exclude the two saturated doublets
since they might have $\logSiIV > 15$, which is the upper
limit we have chosen for comparing \OmSiIV. Including the two
saturated doublets, results in the summed $\OmSiIV > 2.3\times10^{-8}$
for the 18 G = 1 doublets with $\logSiIV \ge 13$.

In order to compare with the high-redshift studies, we assumed the
power-law formalism for \ff{\NSiIV} and integrated the column
density-``weighted'' \ff{\NSiIV} (\ie its first moment):
\begin{equation}
  \OmSiIV = \frac{H_{0}\,\mathrm{m}_{\rm C}} {c\,\rho_{c,0}} 
  \frac{k}{2+\aff{N}}
  \bigg(\frac{\N{\rm max}^{2+\aff{N}} -
    \N{\rm min}^{2+\aff{N}}}{\N{0}^{\aff{N}}} \bigg) 
  {\rm .}  \label{eqn.int_omsiiv}
\end{equation}
From the best-fit values for the G = 1 \ff{\NSiIV}, $\OmSiIV =
(3.71^{+2.82}_{-1.68}) \times 10^{-8}$ for $\log \N{\rm min} = 13$ and
$\log \N{\rm max} = 15$. Since there has been no observed break in
\ff{\NSiIV}, the column density limits are crucial to defining a
finite \OmSiIV\ and comparing between surveys.

The formal errors on the integrated \OmSiIV\ are $2\sigma_{\Omega} =
+5.54/\!-\!2.36$ and $3\sigma_{\Omega} = +8.99/\!-\!2.81$.
derived from the $\aff{N}$ and $\kff{14}$ errors discussed in Section
\ref{subsec.freqdistr}.

\begin{figure*}[!hbt]
  \begin{center}$
    \begin{array}{cc}
      \includegraphics[height=0.47\textwidth,angle=90]{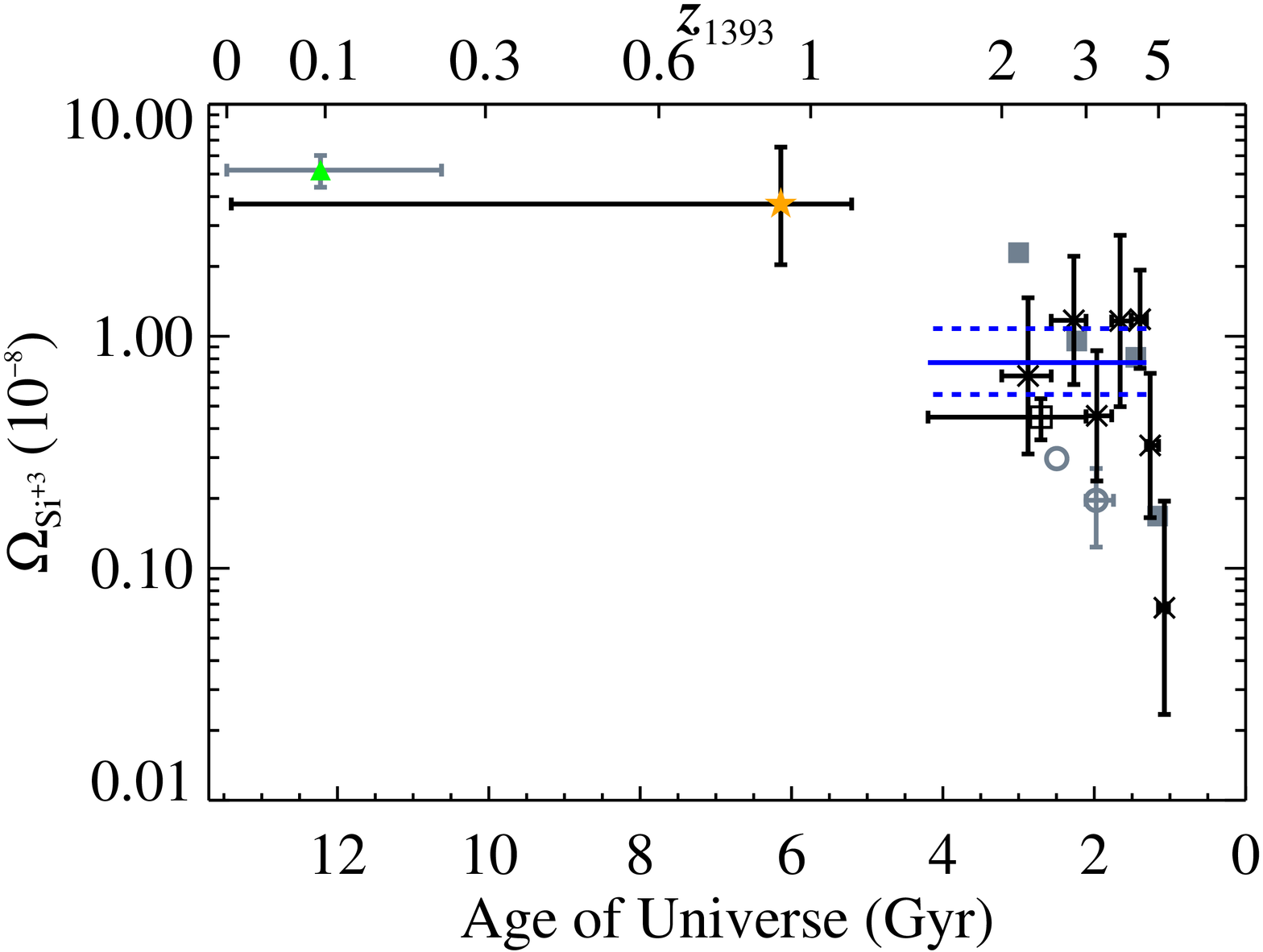} &
      \includegraphics[height=0.47\textwidth,angle=90]{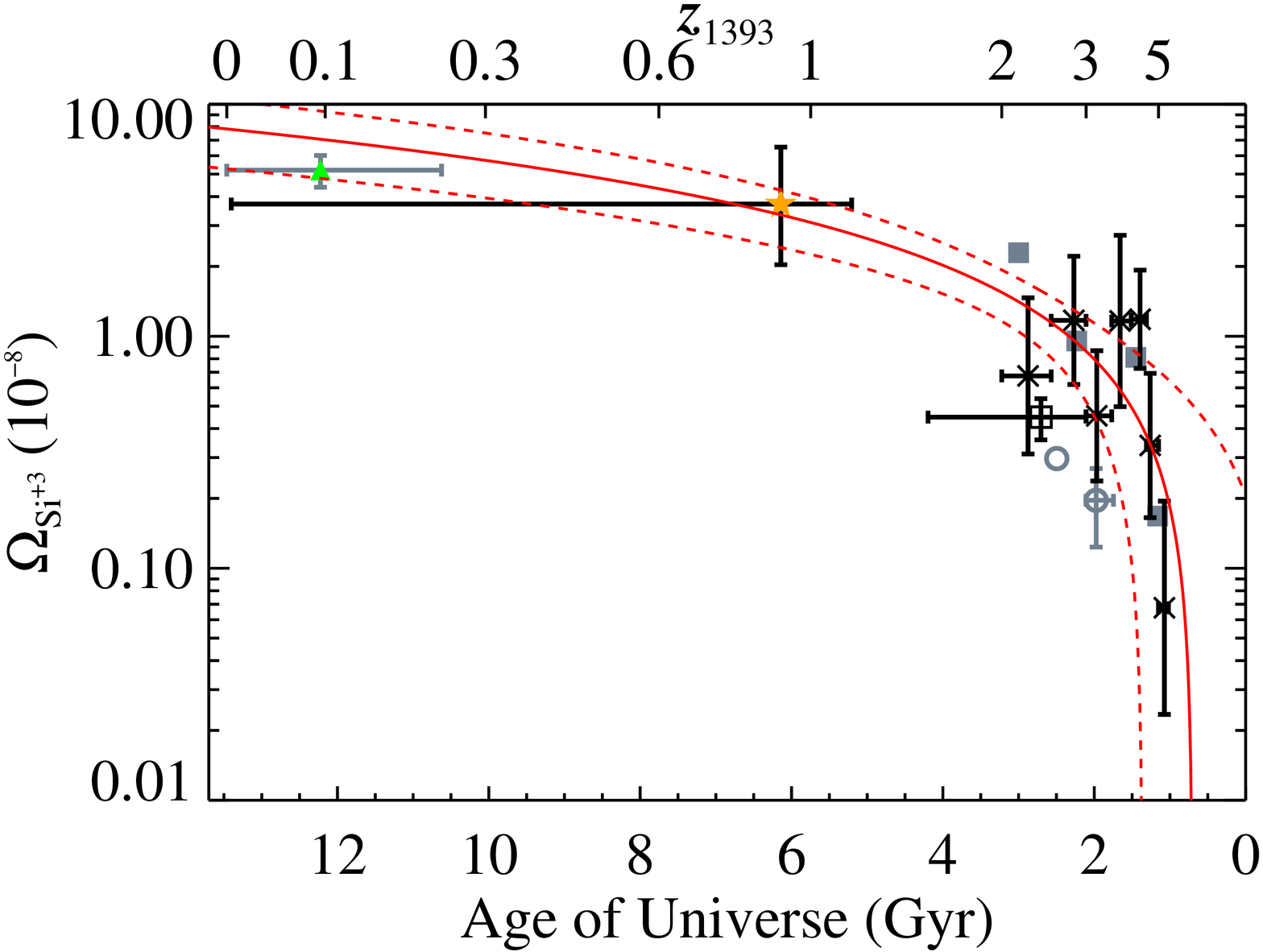}
    \end{array}$
  \end{center}
  \caption[Time evolution of \Sithr\ mass density relative to the critical density.]
  {Time evolution of \Sithr\ mass density relative to the critical
    density. The integrated \OmSiIV, for $13 \le \logSiIV \le 15$ for
    the G = 1 sample, is the (orange) star. The value from
    \citet{danforthandshull08} is the (green) triangle and is
    \emph{not} an independent measurement of \OmSiIV. We compared our
    $z \lesssim 1$ measurement only to those from \citet[][black
    crosses]{songaila01} and \citet[][black, open
    square]{scannapiecoetal06}. The median, and its ``errors'' (see
    Section \ref{subsec.omsiiv}), of the $2 \le z \le 4.5$ values are
    shown on the left with the (blue) solid and dashed lines,
    respectively. The simple linear fit to \OmSiIV\ over $t_{\rm age}$
    is shown on the right by the (red) solid line, where the dashed
    lines are the 1-$\sigma$ range of the fit. For reference, we also
    show the values from \citet[][gray circles]{songaila97} and
    \citet[][gray, filled squares]{songaila05}; the latter of which
    uses a pixel-optical depth method to measure \OmSiIV.
    \label{fig.omsiiv} 
  }
\end{figure*}

We plot the evolution of \OmSiIV\ as a function of the age of the
Universe $t_{\rm age}$ in Figure \ref{fig.omsiiv}.  The
median\footnote{When we quote median values with ``1-$\sigma$
  errors,'' these ``errors'' are actually the difference between the
  median (\ie 50th percentile) and the values at the 15.9th and 84.2nd
  percentiles.}  of the $2 \le z \le 4.5$ studies \citep[][whose
values have been adjusted to match our \NSiIV\ limits and cosmology,
see \citetalias{cookseyetal10}]{songaila01, scannapiecoetal06} is
$\langle \OmSiIV \rangle^{2 \le z \le 4.5} = (0.77^{+0.31}_{-0.21})
\times 10^{-8}$, which is shown by the (blue) lines in Figure
\ref{fig.omsiiv} (left).  We have detected, with $>\!99.8\%$
confidence, an increase in \OmSiIV\ from high-to-low redshift. The $z
\lesssim 1$ is a factor of $4.8^{+3.0}_{-1.9}$ higher than the $2 \le
z \le 4.5$ median.

The confidence limits on the median high-redshift value and the
increase in \OmSiIV\ were estimated based on Monte Carlo sampling of
the distributions. First, we drew $10^6$ realizations of the
high-redshift data sets $\Omega_{\rm MC}^{2 \le z \le 4.5}$, assuming
they had log-normal errors. Then, we measured the median of each set:
$\langle \Omega_{\rm MC}^{2 \le z \le 4.5} \rangle_{i}$. The median
\Sithr\ mass density quoted above ($\langle \OmSiIV \rangle^{2 \le z
  \le 4.5}$) was the median of these ($10^6$) $\langle
\Omega_{\rm MC}^{2 \le z \le 4.5} \rangle_{i}$ values.

To Monte Carlo sample our $z \lesssim 1$ measurement, we used the
likelihood surface discussed in Section \ref{subsec.freqdistr} and the
Metropolis-Hastings algorithm to appropriately sample the $k$-\aff{N}\
parameter space $10^6$ times. For each of these random pairs, we
computed the integrated \Sithr\ mass density (see Equation
\ref{eqn.int_omsiiv}), resulting in a low-redshift Monte Carlo sample:
$\Omega_{\rm MC}^{z \lesssim 1}$. The median of this distribution was:
$\langle \OmSiIV \rangle^{z \lesssim 1} = (3.70^{+1.68}_{-1.19}) \times
10^{-8}$.\footnote{This value is in excellent agreement with our
  integrated value for the best-fit $k$ and \aff{N} and indicates a
  smaller spread, within the quoted confidence limits, than we
  formally adopted (see Table \ref{tab.freqdistr}).} The ratio of the
low- to high-redshift Monte Carlo samples (\ie $\langle \Omega_{\rm
  MC}^{2 \le z \le 4.5} \rangle_{i}/\Omega_{{\rm MC},i}^{z \lesssim 1}$) is
a distribution where the median is $4.8^{+3.0}_{-1.9}$ and the ratio
is greater than unity at the 99.8\% c.l.

A least-squares minimization of a linear model to \OmSiIV\ over
$t_{\rm age}$ for the $2 < z < 5.5$ observations
\citep{songaila01,scannapiecoetal06} and our $z \lesssim 1$ value
yielded: $\ud \OmSiIV / \ud t_{\rm age} = (0.61\pm0.23) \times
10^{-8}\,{\rm Gyr}$, as shown by the (red) lines in Figure
\ref{fig.omsiiv} (right).  This toy-model slope agrees well
($<\!2\sigma$) with $\ud \OmCIV / \ud t_{\rm age} = (0.42\pm0.2)
\times 10^{-8}\,{\rm Gyr}$ from \citetalias{cookseyetal10} for the
equivalent redshift sample but for \ion{C}{4} absorbers, and
\citet{dodoricoetal10} detected a smooth increase in \OmCIV\ from $z =
2.5 \rightarrow 1.5$, which reaches the $z < 1$ \OmCIV\ measured in
\citetalias{cookseyetal10}.  A linear fit to only the $2 \le z < 5.5$
observations indicated no temporal evolution in \OmSiIV\ (\ie $\ud
\OmSiIV / \ud t_{\rm age} = (0.10\pm0.56) \times 10^{-8}\,{\rm Gyr}$).

Modeling \OmSiIV\ as evolving linearly with time was \emph{not} a
physically motivated exercise but one method to evaluate whether our
\OmSiIV\ indicated a significant increase compared to the
high-redshift observations. Since the inclusion of our $z \lesssim 1$
value resulted in a statistically significant rate of increase for
\OmSiIV, we likely have detected a true increase of the \Sithr\ mass
density at low redshift, though proof must await a larger $z \lesssim
1$ survey.

Our results indicate that any increase in \OmSiIV\ at $z \lesssim 1$
is likely due to an increase in the number of high-column density
absorbers (\ie shallower \aff{N} compared to high redshift), since
\dNSiIVdX\ is nearly constant from $z \approx 3 \rightarrow 0$. In
general, the \Sithr\ mass density is dominated by the high-column
density absorbers, so even a small increase in their frequency will
significantly change \OmSiIV.

\subsection{\IRatio}\label{subsec.siivciv}

As mentioned previously, the ionic ratio \IRatio\ has been used to
study the shape and\slash or evolution of the UVB. In order to
construct a complete sample of systems with coverage of both doublets,
we measured the upper limit for \NCIV\ (or \NSiIV) when the doublet
was not detected in association with the targeted \ion{Si}{4} (or
\ion{C}{4}, from \citetalias{cookseyetal10}) doublet but the spectral
coverage existed. Due to how the S/N changed throughout any spectrum, a
\ion{Si}{4}-targeted survey was not sufficient to define a complete
\ion{C}{4} sample and visa versa. Ultimately, there were 12 detections
and 12 lower limits for $z < 1$ \IRatio, with $\logSiIV > 11.9$ and
$\logN(\Cthr) > 13.37$.

\begin{figure*}[!hbt]
  \begin{center}$
    \begin{array}{cc}
      \includegraphics[height=0.47\textwidth,angle=90]{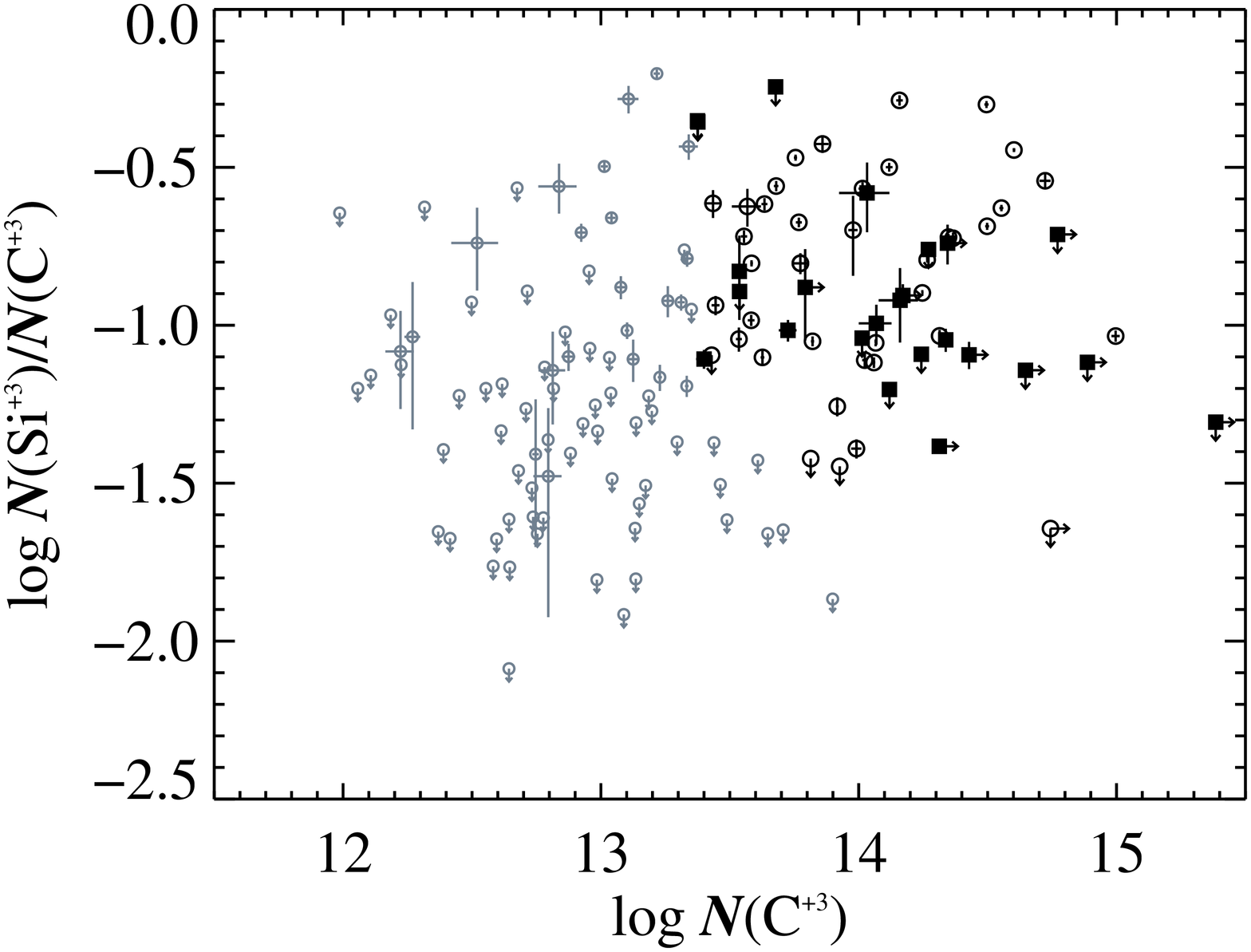} &
      \includegraphics[height=0.47\textwidth,angle=90]{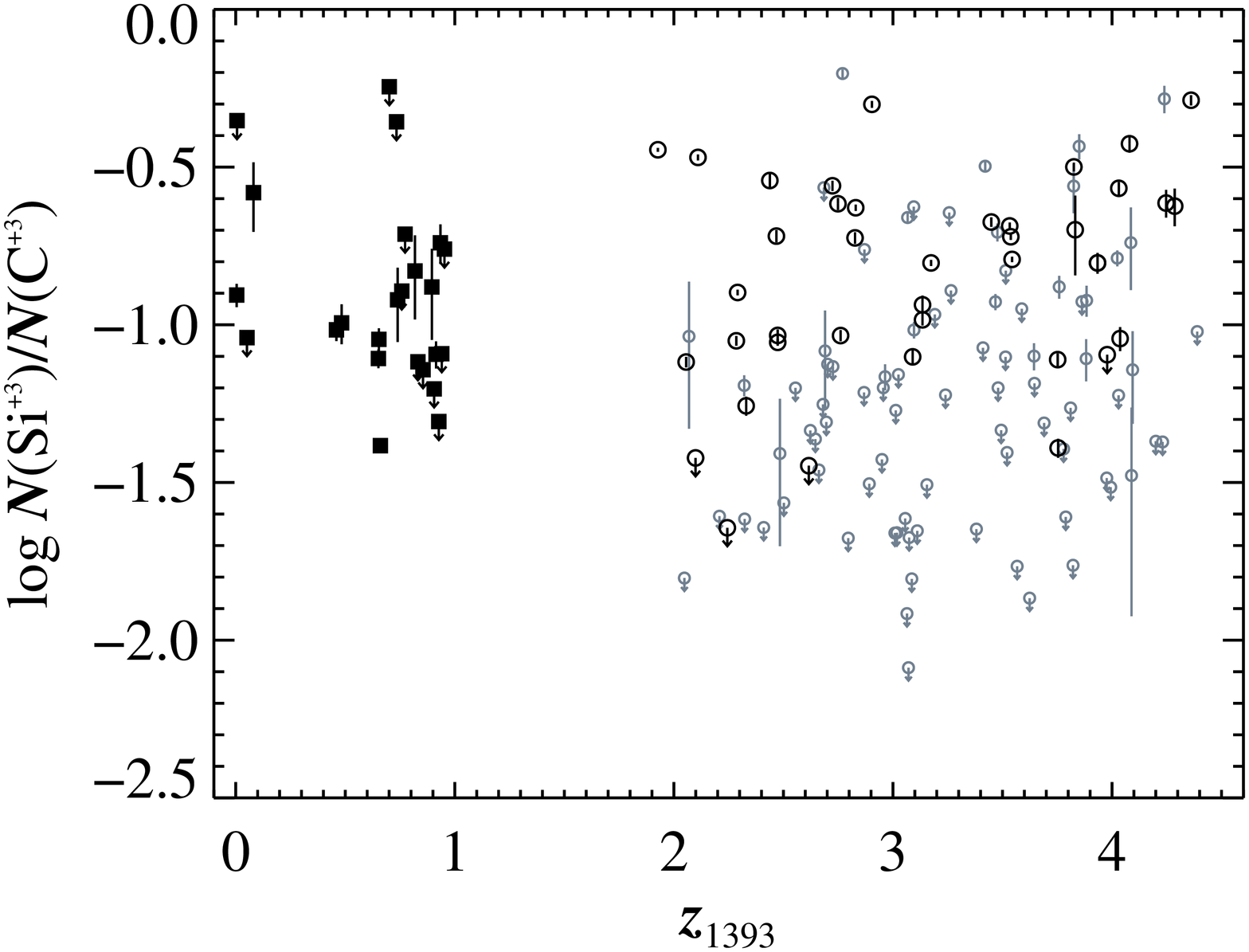}
    \end{array}$
 \end{center}
  \caption[Ionic ratio \IRatio\ as a function of \NCIV\
    and \zsiiv.]
  {Ionic ratio \IRatio\ as a function of \NCIV\
    (left) and \zsiiv\ (right). The $2 \le \zsiiv < 4.5$ data (open
    circles) come from the total column densities per system from
    \citet[][Tables 2--10]{boksenbergetal03ph}. The low-redshift data
    have $\logSiIV > 11.9$ and $\logN(\Cthr) > 13.37$ (black, filled
    squares). High-redshift systems meeting these limits are the larger, black
    circles, while the rest of the sample are the smaller, gray circles. Survival
    analysis shows that the high- and low-redshift ratios are drawn
    from the same parent population and do not evolve significantly
    with redshift.
    \label{fig.siivciv} 
  }
\end{figure*}

In Figure \ref{fig.siivciv}, we compare our \IRatio\ sample with that
from \citet{boksenbergetal03ph}. We reproduced their Figure 16 (bottom
panel) by summing the column densities of all components per system as
given in their Tables 2--10. If there were components with upper
limits for column densities, we set the total system column density to
an upper limit if the components with upper limits were more than 30\%
of the total. If there were components with lower limits for column
densities, we set the total column density to a lower limit. There was
no case when these criteria conflicted. For the high-redshift sample,
there were 39 detections and one upper limit for \IRatio, for doublets
with the observed low-redshift column density limits.

Since both high- and low-redshift samples contained at least one upper
limit, we used survival analysis to enable those limits to contribute
statistically. We used the Astronomy SURVival Analysis package
\citep[ASURV Rev. 1.3, last described in][]{lavalleyetal92} to compare
the two \IRatio\ data sets. First, we tested whether the low-redshift
\IRatio\ distribution shared the same parent population as the
high-redshift ratios. From several univariate ASURV
statistics,\footnote{For more information about univariate analyses
  used here (the two Gehan's, the Peto-Peto, and Peto-Prentice
  generalized tests), see \citet{feigelsonandnelson85}.} we conclude
that the two populations are statistically similar (\ie the null
hypothesis cannot be ruled out with high confidence).

Next we measured the median ratio of the parent population with the
Kaplan-Meier estimator.\footnote{For a useful description of the
  Kaplan-Meier estimator in a context similar to that used here, see
  \citet{simcoeetal04} } The estimated median of the combined low- and
high-redshift data sets was $\langle \IRatio \rangle = 0.16$, and the
25th and 75th percentiles were 0.09 and 0.26, respectively.

Though the estimated means of the high- and low-redshift samples
indicated that there should be no evolution of the ratio with redshift,
we checked for a correlation.\footnote{For more information about the
  bivariate (Cox proportional hazard model, generalized Kendall's tau,
  and Spearman's rho) and linear regression analyses (EM algorithm and
  Buckley-James method) used here, see \citet{isobeetal86}.} Once
again, the null hypothesis (\ie that there is no correlation) cannot
be ruled out with high confidence.

\begin{figure}[!hbt]
  \begin{center}
    \includegraphics[height=0.47\textwidth,angle=90]{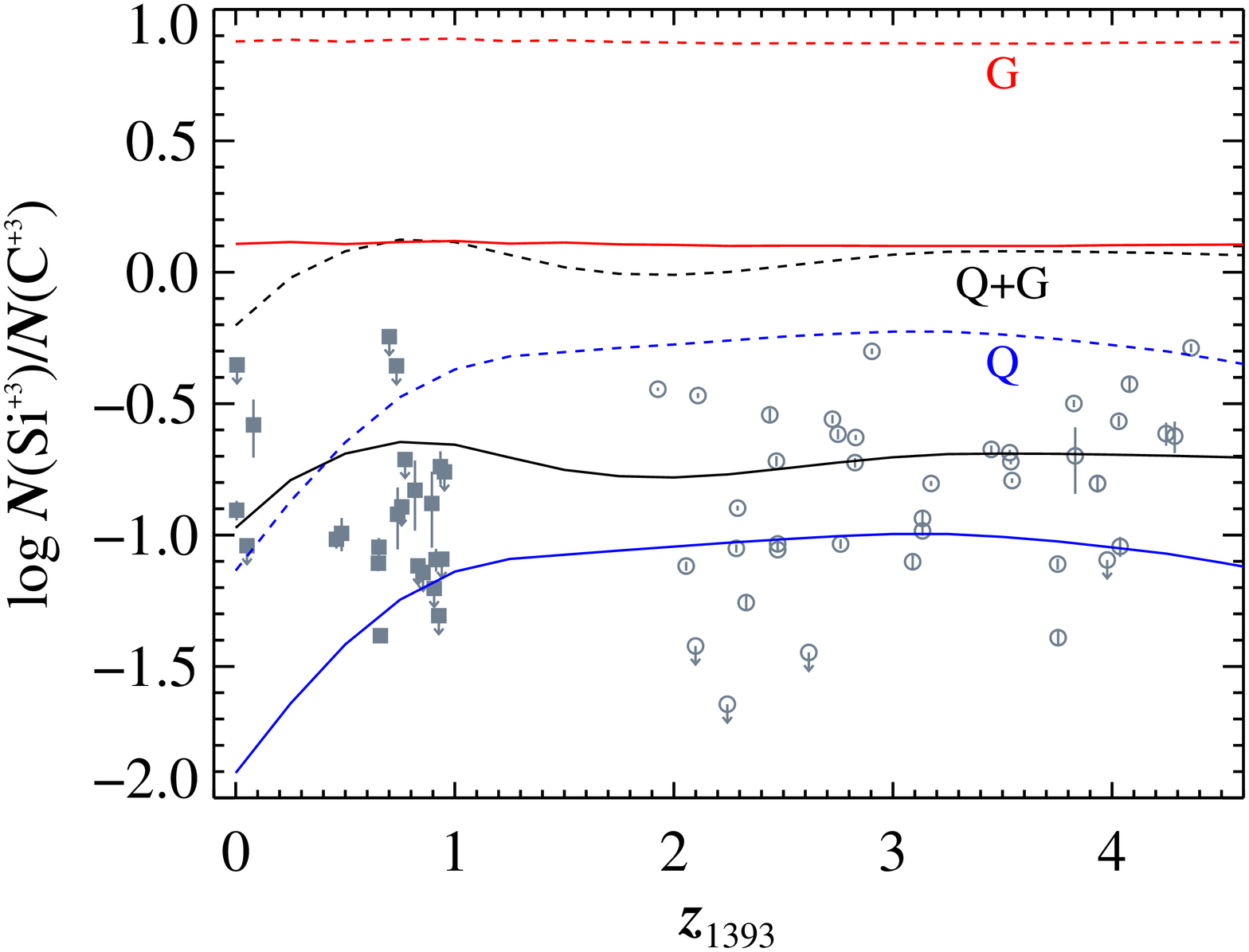}
  \end{center}
  \caption[Predictions from simple photoionization models for \IRatio\
  as a function of \zsiiv.]
  {Predictions from simple photoionization models for \IRatio\ as a
    function of \zsiiv.  All models could reproduce the observed lack of
    evolution in \IRatio\ with redshift and the overall magnitude of
    \IRatio\ through careful choice of the CLOUDY parameters (\eg
    $\log U$, [Si/C]). For this plot, the ionization parameter is
    $\log U = -2$; the metallicity is $0.001\,Z_{\odot}$. The solid
    lines indicate models with solar relative abundances and the
    dashed lines, Si-enhanced (\ie $[{\rm Si}/{\rm C}] = +0.77$).  The
    gray, background points are the observations (black symbols from
    Figure \ref{fig.siivciv}, right), but the CLOUDY models were
    \emph{not} fit to the observations.
    \label{fig.cloudy} 
  }
\end{figure}

\subsection{Nature of Systems with \ion{Si}{4} and
  \ion{C}{4} Absorption}\label{subsec.nature} 

We concluded that there has been no evolution in \IRatio\ from $z
= 4.5 \rightarrow 0$, based on our sample and that from
\citet{boksenbergetal03ph}. Next we explored what the lack of evolution
means, and we began by disentangling the physics involved in the ratio
\IRatio:
\begin{equation}
  \frac{\NSiIV}{\NCIV} = 
   \Bigg( \frac{\displaystyle L_{\rm Si}}{\displaystyle L_{\rm C}}
  \Bigg)  \Bigg( \frac{\displaystyle n_{\rm Si}}{\displaystyle n_{\rm C}} \Bigg)
  \Bigg( \frac{\displaystyle \chi^{\Sithr}_{\rm Si}}{\displaystyle
    \chi^{\Cthr}_{\rm C}} \Bigg)
  {\rm ,}
  \label{eqn.siivciv}
\end{equation}
where $L_{\rm X}$ is the size of the cloud (enriched with X), along
the line of sight; $n_{\rm X}$ is the volume density of element X; and
$\chi^{\rm X'}_{\rm X}$ is the fraction of X ionized into ion ${\rm
  X'}$.  The first term on the right-hand side is affected by the
spatial distribution; the second, the metal abundances; and the third,
the ionizing background.

The Universal trend is for structure to collapse and become denser
with age in a $\Lambda$CDM Universe and for feedback processes to
disperse and mix metals on varying scales. Therefore, more of the
filamentary structure is enriched as the Universe ages; though
feedback may preferentially enrich voids instead of filaments, as a
result of the density difference \citep{kawataandrauch07}. However, in
our systems with both \ion{Si}{4} and \ion{C}{4} absorption, the
absorption profiles trace each other quite well. There are no obvious
system where \IRatio\ varies significantly between the
components. Therefore, we infer $L_{\rm Si} \approx L_{\rm C}$.

We know that the metallicity of the Universe increases with age, but
the relative abundance of silicon and carbon does not, necessarily,
follow suit. Hence, the evolution in $n_{\rm Si}/n_{\rm C}$ is unclear.

We explored the effect of the ionizing background with a suite of
simple photoionization models, in order to develop our understanding
of the last term in Equation \ref{eqn.siivciv}. We used the spectral
synthesis program CLOUDY v08, as last described by
\citet{ferlandetal98}.  We modeled the medium as a plane-parallel
slab, ionized by the \citet[][updated in 2005]{haardtandmadau96}
ultraviolet background, for quasar (Q), galaxy (G), and quasar+galaxy
(Q+G) models. We set the number density of hydrogen $n_{\rm H} =
0.1\cm{-3}$, though our models were insensitive to this parameter in
the optically-thin regime. We assumed a neutral column density $\log
\N{HI} = 16$ and metallicity $Z = 0.001\,Z_{\odot}$.  We tested two
cases: solar relative abundances and Si-enhanced; the increase in silicon was
such that $[{\rm Si/C}] = +0.77$, as measured in
\citet{aguirreetal04}. We varied the ionization parameter $\log U =
[-5,-4,-3,-2,-1]$, which is a dimensionless ratio of the flux of
hydrogen-ionizing photons to $n_{\rm H} \cdot c$.

All models could reproduce the observed lack of evolution in the ionic
ratios with redshift.  In Figure \ref{fig.cloudy}, we show \IRatio\ as
a function of redshift for the three UVB models, with solar relative
abundances and Si-enhanced, and for $\log U = -2$. We did \emph{not}
fit the observations, but they are reproduced well by the solar
relative abundance, Q+G model. More importantly, all UVB models could
reproduce a shape consistent with the observed lack of evolution in
\IRatio, given the right choices of \eg $\log U$, $\log \N{HI}$,
[Si/C]. The overall magnitude is nearly freely scalable by adjusting
these parameters, because the observed sample of \ion{Si}{4} and
\ion{C}{4} doublets are \emph{not} drawn from a single type of cloud,
as we modeled. Evidently, there is no need for a particularly soft UVB
(\ie model G) to reproduce the lack of redshift evolution in \IRatio.

For there to be no evolution in \IRatio\ from $z = 4.5 \rightarrow 0$,
the abundance, ionizing background, and structure of silicon- and
carbon-enriched gas are constrained to be in ``balance.'' The
observations indicate that these three processes evolved to maintain a
constant ratio of $\langle \IRatio \rangle = 0.16$ for nearly 12\,Gyr,
for absorbers with $\logSiIV > 11.9$ and $\logN(\Cthr) > 13.37$.

Disentangling the effect of the detailed physics (\eg changing silicon
and carbon abundances, variation in the physical properties of
absorbing clouds) require cosmological hydrodynamic simulations that
could resolve enrichment processes in galactic halos and the large-scale
structure.

\section{Summary}\label{sec.summ}

We conducted a blind survey for $z \lesssim 1$ \ion{Si}{4} doublets in
the \hst\ UV spectra of 49 quasars. We identified 22 definite
\ion{Si}{4} systems (G = 1) and six ``high-likely'' ones (G = 2), and
this represents the largest sample of low-redshift \ion{Si}{4}
doublets prior to Servicing Mission 4.  From a sample of 20 $z
\lesssim 1$ \ion{Si}{4} doublets with both lines detected at $\EWr \ge
3\sigEWr$, we measured a line density $\dNSiIVdX = 1.4^{+0.4}_{-0.3}$
for $\logSiIV > 12.9$.

We constructed frequency distributions of the column densities
\ff{\NSiIV}\ and the rest equivalent widths \ff{\EWlin{1393}}. Both
were approximated well by power laws. The
best-fit power law to \ff{\NSiIV}\ had slope $\aff{N} =
-1.61^{+0.28}_{-0.31}$ and normalization $k = (1.18^{+0.45}_{-0.36})
\times 10^{-14}\cm{2}$.  We compared the \ion{Si}{4} line density to
high-redshift observations by integrating \ff{\NSiIV}, and \dNSiIVdX\
does not evolve significantly from $z \approx 3 \rightarrow 0$.

From the first moment of \ff{\NSiIV}, we measured the \Sithr\ mass
density relative to the critical density: $\OmSiIV =
(3.71^{+2.82}_{-1.68}) \times 10^{-8}$ for $13 \le \logSiIV \le
15$. This value was estimated, with Monte Carlo sampling of the
distributions, to be a factor of $4.8^{+3.0}_{-1.9}$ greater than the
measurements from the $2 \le z \le 4.5$ studies from
\citet{songaila01} and \citet{scannapiecoetal06}.

From a simple linear fit, we estimated the rate of increase in the
\Sithr\ mass density over time to be $\ud \OmSiIV / \ud t_{\rm age} =
(0.61\pm0.23) \times 10^{-8}\,{\rm Gyr}$. Though a linear model is
extremely simplistic and not physically motivated, it does lend
support to the $z \lesssim 1$ \OmSiIV\ being a true increase over the
$1.5 < z < 5.5$ observations, which, when fit by themselves, result in
no statistically significant temporal evolution.

Any increase in \OmSiIV\ is probably driven by the increase in the
number of high-column density absorbers (\ie shallower \aff{N}), since
\dNSiIVdX\ does not increase significantly from $z \approx 3
\rightarrow 0$.

We also compared the ionic ratio \IRatio\ from the current study and
\citetalias{cookseyetal10} with the high-redshift sample of
\citet{boksenbergetal03ph}. From survival analysis of the two
populations, we concluded that the ionic ratios of the high- and
low-redshift distributions are drawn from the same parent population,
with median $\langle \IRatio \rangle = 0.16$. The lack of evolution in
\IRatio\ from $z = 4.5 \rightarrow 0$ places constraints on the
evolution of the metal production, feedback processes, and the
ionizing background. These three processes evolve in some balanced
fashion to have \ion{Si}{4} and \ion{C}{4} absorbers evolve in
lock-step for the last $\approx\!12\,{\rm Gyr}$.

We explored the effect of the ionizing background on the
(non)evolution of \IRatio\ with a suite of simple CLOUDY models. We
varied the background by using the canonical \citet{haardtandmadau96}
quasar, galaxy, and quasar+galaxy models. All three backgrounds could
result in relatively constant \IRatio\ over redshift, for models with
the ``right'' choices of \eg ionizing parameter, [Si/C].  Therefore,
a soft UVB (\ie model G) is not preferred.

In general, more observations---at low and high redshift---are needed
to increase the statistical significance of the trends that are
currently highly suggestive. We eagerly anticipate new low-redshift
results from the \hst\ Cosmic Origins Spectrograph
\citep[COS,][]{morseetal98}. Meanwhile, cosmological hydrodynamic
simulations should be leveraged to understand how metal production and
dispersal and the ionizing background interact to evolve \ion{Si}{4}
and \ion{C}{4} absorbers in tandem.


\acknowledgements We would like to thank P. Jonsson, for helping with
the Monte Carlo analyses. We also thank the anonymous referee, for
constructive comments that helped improve the paper.  This study is
based on observations made with the NASA-CNES-CSA \emph{Far
  Ultraviolet Spectroscopic Explorer}. \fuse\ is operated for NASA by
the Johns Hopkins University under NASA contract NAS5-32985.  This
work is also based on observations made with the NASA/ESA \emph{Hubble
  Space Telescope} Space Telescope Imaging Spectrograph and Goddard
High-Resolution Spectrograph, obtained from the data archive at the
Space Telescope Institute. STScI is operated by the association of
Universities for Research in Astronomy, Inc. under the NASA contract
NAS 5-26555.  The current study was funded by the HST archival grant
10679; the NSF CAREER grant AST 05\_48180; and the MIT Department of
Physics.

{\it Facilities:} \facility{\fuse}, \facility{\hst (\stis)},
\facility{\hst (\ghrs)}

\bibliographystyle{apj} \bibliography{../../civabsorbers}

\end{document}